\documentclass[aps,prx,superscriptaddress,twocolumn,amsmath,amssymb]{revtex4-2}

\usepackage{bm}% bold math
\usepackage{graphicx}% Include figure files
\usepackage{hyperref}% add hypertext capabilities
\hypersetup{hypertex=true,
            colorlinks=true,
            linkcolor=red,
            urlcolor=black,
            citecolor=blue}

\begin{document}

%\preprint{}

\title{Localized elasticity governs the nonlinear rheology of colloidal supercooled liquids}

\author{Dejia Kong}
\affiliation{Department of Engineering Physics and Key Laboratory of Particle and Radiation Imaging (Tsinghua University) of Ministry of Education, Tsinghua University, Beijing 100084, China}

\author{Wei-Ren Chen} 
\affiliation{Neutron Scattering Division, Oak Ridge National Laboratory, Oak Ridge, Tennessee 37831, United States}

\author{Ke-Qi Zeng}
\affiliation{Department of Engineering Physics and Key Laboratory of Particle and Radiation Imaging (Tsinghua University) of Ministry of Education, Tsinghua University, Beijing 100084, China}

\author{Lionel Porcar}
\affiliation{Institut Laue-Langevin, B.P. 156, F-38042 Grenoble CEDEX 9, France}
 
\author{Zhe Wang}
\email[Corresponding author: ]{zwang2017@mail.tsinghua.edu.cn}
\affiliation{Department of Engineering Physics and Key Laboratory of Particle and Radiation Imaging (Tsinghua University) of Ministry of Education, Tsinghua University, Beijing 100084, China}

\date{\today}

\begin{abstract}
We propose a microscopic picture for understanding the nonlinear rheology of supercooled liquids with soft repulsive potentials. Based on Brownian dynamics simulations of supercooled charge-stabilized colloidal suspensions, our analysis shows that the shear thinning of viscosity ($\eta$) at large enough shear rates ($\dot{\gamma}$), expressed as $\eta\sim\dot{\gamma}^{-\lambda}$, originates from the evolution of the localized elastic region (LER). An LER is a transient zone composed of the first several coordination shells of a reference particle. In response to the external shear, particles within LER undergo nearly affine displacement before the yielding of LER. The characteristic strain ($\gamma$) and size ($\xi$) of LER respectively depend on the shear rate by $\gamma\sim\dot{\gamma}^\epsilon$ and $\xi\sim\dot{\gamma}^{-\nu}$. Three exponents, $\lambda$, $\epsilon$, and $\nu$, are related by $\lambda=1-\epsilon=4\nu$. This simple relation connects the nonlinear rheology to the elastic properties and the microscopic configurational distortion of the system. The relaxation of the LER is promoted by the large-step nonaffine particle displacement along the extensional direction of the shear geometry with the step length of 0.4 particle diameter. The elastic deformation and relaxation of the LER are ubiquitous and successive in the flow, which compose the fundamental process governing the bulk nonlinear viscoelasticity. We apply this model to analyze the Rheo-Small Angle Neutron Scattering data of sheared charge-stabilized colloidal suspensions. It is seen that our model well explains the neutron spectra and the rheological data.
\end{abstract}

%\keywords{}

\maketitle

\section{INTRODUCTION}

The flow of supercooled liquids and glassy materials is common in nature, daily life, and a variety of industrial fields. Therefore, understanding this phenomenon is of fundamental and practical importance \cite{larson1}. With extensive theoretical and numerical investigations \cite{spaepen1, argon1, langer1, lemaitre1, barrat0, voigtmann1} and experimental evidences from colloidal glasses \cite{weitz1, poon1, weeks1}, now it is known that for glassy solids, the shear-induced yielding and flow stem from the formation and accumulation of the “shear transformation zone” (STZ), which contains a few of particles undergoing irreversible nonaffine displacements. The localized plastic events are found to exhibit strain-rate-dependent long-range anisotropic correlations \cite{weitz1, schall1, schall2, schall3, picard1, maloney1, barrat1}. Such correlations facilitate the formation of new STZs in the vicinities of existing ones \cite{weitz1} and, under certain circumstances, result in inhomogeneous flow \cite{schall1, lemaitre2, tsamados1}. The size of the STZ is linked to the particle self-diffusion, which bridges the rheological behaviors and the dynamics on the particle level \cite{egelhaaf1, barrat2}. An alternative approach to understand the rheology of glasses is to focus on the behaviors of the “cage”, formed by the nearest neighbors of a reference particle \cite{pete1, poon2, pete2, pete3, pete4, egelhaaf2}. With rheological \cite{pete1, poon2, pete2, pete3, pete4}, small-angle scattering \cite{schall4, schall5, fuchs1} and confocal measurements \cite{pete4} complemented by Brownian Dynamics (BD) simulation \cite{pete2, pete4, brady1}, the macroscopic deformation and yielding of the colloidal glass are respectively related to the elastic deformation and rearrangement of the cage. The caging and the shear-driven cage-breaking effects tie in well with the extension of the mode-coupling theory (MCT) \cite{gotze1} into the flowing glasses \cite{fuchs2, fuchs3, fuchs4}. Further studies of metallic glasses by X-ray diffraction \cite{egami1} and computer simulation \cite{egami2} suggest that the spatial range of the local elastic deformation in flowing glasses extends beyond the size of the cage.

On the other side of the glass transition point, materials may fail to crystallize and stay as liquid. When subject to slow shear, these supercooled liquids are free to flow but are enormously viscous compared to the normal liquids \cite{cavagna1, berthier1}. Upon increasing the shear rate $\dot{\gamma}$, the viscosity $\eta$ of supercooled liquids becomes shear thinning \cite{hoover1, simmons1, miyazaki1} and progressively approaches a relation of $\eta\sim\dot{\gamma}^{-\lambda}$ with $\lambda\le1$ \cite{miyazaki1, yama1, yama2}. Compared with the vast literature on the flow of glasses, less attention has been paid to the nonlinear rheology of supercooled liquids. One of the current understandings on the anomalous viscosity of the flowing supercooled liquids is built on the concept of “dynamical heterogeneity”, i.e., temporary regions where particles cooperatively undergo large displacements to realize structural rearrangements \cite{weitz1, cavagna1, yama1, yama2, ediger1, adam1, glotzer1, harrowell1, yama3, tanaka1, lemaitre3}. Previous numerical studies have revealed the existence of temporary clusters of bond breaks in sheared atomic supercooled liquids \cite{yama1, yama2}. Similar dynamically heterogeneous effects were also characterized by various four-point correlation functions \cite{glotzer1, yama3, tanaka1}. These studies have established the correspondence between the evolution of the properties of the dynamical clusters, such as the size and life time, and the shear thinning of supercooled liquids \cite{yama2, yama3, tanaka1}. Interestingly, while many simulation results show that the supercooled liquid becomes more dynamically homogeneous as shear rate increases \cite{yama1, yama2, yama3}, it is suggested that in the shear-thinning regime the mobile regions tend to form anisotropic fluidized bands \cite{tanaka1}. On the other hand, MCT framework provides a homogeneous description of the nonlinear rheology of supercooled liquids close to the glass transition by encoding the caging effect and the resulting non-Markovian slow dynamics of density fluctuations \cite{fuchs2}. Its prediction nicely agrees with the experimental data of hard-sphere-like microgels \cite{fuchs3, fuchs4}. Despite these efforts, the structural indicator, through which the shear thinning behavior can be directly connected to the microscopic distortion in flowing supercooled liquids, remains elusive. This topic is the focus of this work.

Herein, we adopt concentrated colloidal suspensions as the model system to explore the flowing state of supercooled liquids. Colloidal systems possess large particle size that enables to probe the particle-level structure and dynamics through confocal microscopy and scattering methods \cite{schall5, weeks2, weitz2, ackerson1}. The inter-colloid interaction can be well controlled and characterized \cite{wagner1, brader1, vla1, pusey1}, which adds to the flexibility in experimental studies. Moreover, by introducing the effective interparticle interaction, slow dynamics and collective phase behaviors of colloidal suspensions can be mapped onto those of atomic liquids or other condensed systems \cite{voigtmann2, sciortino1, weitz3, pusey2, noyola1, wang0}. Hence, colloidal suspensions have been extensively used for experimentally verifying the theoretical and numerical predictions built on atomic liquids \cite{poon3, poon4}. An important feature distinguishing colloidal suspensions from atomic liquids is that the colloids can interact via the hydrodynamic force resulting from the motion of solvent \cite{dhont1, brady2, wagner2}. At high enough shear rates and concentrations, it is suggested that hydrodynamic lubrication forces spawn the hydroclusters, which cause the shear thickening phenomenon that does not exist in atomic systems \cite{brady2, wagner2, wagner3, wagner4}. More recent studies show that the frictional contact between colloidal particles plays the crucial role in inducing the shear thickening at high concentrations \cite{isa1,seto1,cates0}. The shear thickening effect is particularly common in the most widely-used hard-sphere colloids \cite{pusey1}. To suppress this effect and highlight the shear thinning behavior, we adopt the charged-stabilized colloidal suspensions with long-range screening Coulomb potential \cite{nagele1}. Firstly, the Coulomb repulsion inhibits the shear thickening by weakening the near-contact lubrication effect and preventing the contact \cite{morris1}. Secondly, theoretical and experimental studies \cite{morris1,russel1} prove that the Coulomb repulsion remarkably enhances the shear thinning. From the viewpoint of energy, the strain-induced distortion and rearrangement of local configuration respectively stores and depletes the elastic free energy arising from the Coulomb potential, which contribute to the bulk viscoelasticity \cite{lacks1, lacks2, lacks3}. While this mechanism does not exist in hard-sphere colloids \cite{cohen1}. Considering these facts, we suggest that using charged colloids rather than hard-sphere colloids makes it more convenient to compare with the previous literature on atomic liquids with soft repulsive potentials \cite{yama1, yama2, yama3, tanaka1, lemaitre3} and to reveal the role of the long-range interparticle repulsion in the nonlinear rheology.

For flowing liquids, the average change of the local configuration is reflected by the distortion of the pair distribution function $g(\bm{r})$ \cite{dhont1, morris1, wagner5} defined as $\rho g(\bm{r})=\langle\sum_{i=2}^{N}\delta[\bm{r}-(\bm{r}_i-\bm{r}_1)]\rangle$, where $\bm{r}_i$ is the position of particle $i$, $\bm{r}$ denotes the displacement from the reference particle located at $\bm{r}_1$, $N$ is the particle number, $\rho$ is the average particle number density, and $\langle\dots\rangle$ denotes the thermal average \cite{hansen1}. Inspired by the idea of local elasticity in flowing glasses \cite{pete1, pete4, egami2, cates1}, we performed a systematic investigation on the $g(\bm{r})$ of sheared supercooled liquids generated by BD simulations. The results suggest that the elastic deformation and relaxation of the \textit{localized elastic region} compose the microscopic source of the nonlinear rheology of the system. In response to the external shear, such a region, which contains the first several coordination shells of a reference particle, undergoes solid-like deformation until its yielding. The length scale of the region shrinks with the shear rate as $\dot\gamma^{-\nu}$, while the characteristic strain of the region is enhanced by shear as $\dot\gamma^\epsilon$. The exponents, $\epsilon$, $\nu$, and $\lambda$ that appears in $\eta\sim\dot\gamma^{-\lambda}$, are related by $\lambda=4\nu=1-\epsilon$. The relaxation of the localized elastic region is mainly promoted by the large-step nonaffine particle displacement along the extensional direction of the shear geometry. The step length is about 0.4 particle diameter. Such nonaffinity becomes more prominent upon increasing the shear rate. We also performed Rheo-Small Angle Neutron Scattering (Rheo-SANS) \cite{porcar1, hel1, burgh1} experiments on concentrated charge-stabilized colloidal suspensions under steady shear. The experimental result proves that the local elasticity plays a dominant role in the shear thinning behavior. Moreover, we find the clue of the shrinkage of the LER with shear rate. These observations are well consistent with the predictions of our model.

The rest of the paper is organized as follows. In section II, we present a framework for analyzing the distorted $g(\bm{r})$. An empirical kinetic equation is introduced. Spherical harmonic expansion method is employed to extract the most relevant information from $g(\bm{r})$. In section III we identify the localized elastic region in sheared supercooled liquids from the BD results. Relevant length scale and properties are explored in III.A, and the yielding of this transient localized elasticity is discussed in III.B. Section IV provides the details and results of our Rheo-SANS experiments. Concluding remarks are included in section V.

\section{THEORETICAL FRAMEWORK}

Smoluchowski equation is a common point of departure for the theoretical investigation on colloidal dynamics \cite{dhont1, morris2, russel2}. While in scattering and simulation studies, empirical kinetic equations were widely employed to analyze the shear-induced microstructural anisotropy because of their simplicity \cite{hess1, hess2, hess3, hess4, ackerson2}. For a liquid undergoing shear flow with the stream velocity along the $x$-direction, the velocity gradient along the $y$-direction, and the shear rate $\dot\gamma$,the kinetic equation of the pair correlation $g(\bm{r},t)$ can be written as \cite{hess2, hess3}
\begin{equation}
\frac{\partial}{\partial t}g(\bm{r},t)+\dot\gamma y\frac{\partial}{\partial x}g(\bm{r},t)+\Omega(g)=0.\label{eq:2_1}
\end{equation}
In Eq.~(\ref{eq:2_1}), the second term represents the convective distortion, and $\Omega(g)$ denotes the damping effect on this distortion. Generally, $\Omega(g)$ depends on the Brownian effect, particle distribution and interparticle interaction. A typical form of $\Omega(g)$ is given by $\Omega(g)=-2D_0\nabla\cdot\{\nabla g(\bm{r},t)-[\nabla\ln g_\text{eq}(r)]g(\bm{r},t)\}$, where $D_0$ is the Stokes-Einstein self-diffusion coefficient and $g_\text{eq}(r)$ is the pair distribution function at zero shear. This choice of $\Omega(g)$ leads to the Smoluchowski equation of pair correlation, which can be obtained by integrating over unnecessary degrees of freedom in the $N$-body Smoluchowski equation and neglecting hydrodynamic interaction \cite{dhont1}. In this work, a practical $\Omega(g)$ is adopted. Considering that $\Omega(g=g_\text{eq})=0$ at the quiescent state, an acceptable form of $\Omega(g)$ for the fluid under steady shear can be written as
\begin{equation}
\Omega(g)=\tau^{-1}(\bm{r})[g(\bm{r})-g_\text{eq}(r)],\label{eq:2_2}
\end{equation}
where $\tau(\bm{r})$ has a dimension of time.

Spherical harmonic expansion (SHE) is a frequently-used method for analyzing the anisotropy of particle distribution in flowing fluids \cite{fuchs1, egami2, hess3, hess4, ackerson2, chen1} and deformed polymers \cite{wang1, wang2, weig1}. This approach allows a convenient extraction of the key information that bridges the microscopic distortion and the rheological behavior according to the deformation geometry \cite{egami2, hess3, wang2}. Here, we perform the SHE on $g(\bm{r})$, and the following expansion is found
\begin{equation}
g(\bm{r})=\sum_{l=0}^{\infty}\sum_{m=-l}^lg_l^m(r)Y_l^m(\bm{\Omega}),\label{eq:2_3}
\end{equation}
where $g_l^m(r)$ is the expansion coefficient, and $Y_l^m(\bm{\Omega}=\bm{r}/r)$ is the tesseral (real basis) spherical harmonic function defined as
\begin{widetext}
\begin{equation}
Y_l^m(\bm{\Omega})=Y_l^m(\theta,\phi)=
\left\{ \begin{aligned} 
& \sqrt{2}\sqrt{(2l+1)\frac{(l-|m|)!}{(l+|m|)!}}P_l^{|m|}(\cos\theta)\sin{(|m|\phi)} & (m<0) \\
& \sqrt{2l+1}P_l^0(\cos\theta) & (m=0) \\
& \sqrt{2}\sqrt{(2l+1)\frac{(l-m)!}{(l+m)!}}P_l^{m}(\cos\theta)\cos{(m\phi)} & (m>0)
\end{aligned} \right. ,\label{eq:2_4}
\end{equation}
\end{widetext}
where $P_l^m(x)$ is the associated Legendre polynomial, $\theta$ is the polar angle from the positive $z$ axis, and $\phi$ is the azimuthal angle in the $x-y$ plane from the positive $x$ axis. The feature of the expansion given in Eq.~(\ref{eq:2_3}) is determined by the shear geometry. Due to the symmetry of $g(r,\theta,\phi)=g(r,\pi-\theta,\phi)$ and $g(r,\theta,\phi)=g(r,\theta,\phi+\pi)$, only terms with even $l$ and $m$ survive. In addition, it is seen that the pattern of $Y_2^{-2}(\theta,\phi)$ ($Y_2^{-2}\propto\sin^2\theta\sin{2\phi}\propto\hat{x}\hat{y}$) is consistent with the shear geometry. Thus, $g_2^{-2}(r)$ should be the most prominent anisotropic term \cite{hoover1, hess3, hess4}. As $l$ further increases, the pattern of $Y_l^m$ becomes more and more complicated, and the magnitude of $g_l^m(r)$ is expected to progressively weaken.

$\Omega(g)$ can also be expressed by spherical harmonics. Based on computer simulation results, Hess \textit{et al}. suggest that anisotropic terms with the same $l$ correspond to similar characteristic relaxation time \cite{hess2, hess3, hess4}. Consequently, $\Omega(g)$ is approximated as
\begin{equation}
\begin{aligned}
\Omega(g)&\approx\tau_0^{-1}(r)[g_0^0(r)-g_{\text{eq}}(r)]\\
&+\sum_{l=2}\tau_l^{-1}(r)\sum_{m=-l}^lg_l^m(r)Y_l^m(\bm\Omega).\label{eq:2_5}
\end{aligned}
\end{equation}
Inserting Eq.~(\ref{eq:2_5}) into Eq.~(\ref{eq:2_1}) yields a group of coupled equations for $g_l^m(r)$. To decouple these equations, one can expand them with respect to $\dot\gamma\tau_l$ in the case of $\dot\gamma\tau_l<1$, and terminate the expansions at certain order of $\dot\gamma\tau_l$. To the first order of $\dot\gamma\tau_l$, one has
\begin{equation}
g_2^{-2}(r)=-\frac{1}{\sqrt{15}}\dot\gamma\tau_2(r)r\frac{\textrm{d}}{\textrm{d}r}g_0^0(r).\label{eq:2_6}
\end{equation}
To second order of $\dot\gamma\tau_l$, one obtains the relations involves $g_2^0$, $g_2^{-2}$, $g_4^0$ and $g_4^4$. Currently, we focus on the first-order result. Equation (\ref{eq:2_6}) links the two most prominent terms, the isotropic term $g_0^0(r)$ and the leading anisotropic term $g_2^{-2}(r)$. It plays a central role in our following analysis. By assuming $\tau_2(r)$ as a constant about $r$, equations similar to Eq.~(\ref{eq:2_6}) have been widely used to quantify the shear-induced microscopic anisotropy in atomic liquids \cite{hoover1, hess2, hess3, ronis1} and colloidal suspensions \cite{ackerson2, russel3}. In principle, the functional form of $\tau_2(r)$ depends on the shear rate, the concentration of colloidal particle, and the interparticle interaction \cite{batchelor1}. In the following parts, we will show that the form of $\tau_2(r)$ is closely related to the way of the response of the liquid to the imposed shear.

Since we are going to explore the viscoelasticity of the sheared liquids, it could be useful to review the microscopic anisotropy induced by elastic deformation. For a solid that undergoes an affine deformation with a small shear strain $\gamma$, it is straightforward to find that, to the first order of $\gamma$, $g_2^{-2}(r)$ is written as \cite{egami3}
\begin{equation}
\begin{aligned}
g_2^{-2}(r)&\approx-\frac{1}{\sqrt{15}}\gamma r\frac{\textrm{d}}{\textrm{d}r}g_{\text{eq}}(r)\\
&\approx-\frac{1}{\sqrt{15}}\gamma r\frac{\textrm{d}}{\textrm{d}r}g_0^0(r).\label{eq:2_7}
\end{aligned}
\end{equation}
This equation works very well for $\gamma\approx0.1$ or less. To the first order of $\gamma$, $g_{\text{eq}}(r)$ and $g_0^0(r)$ in Eq.~(\ref{eq:2_7}) can be replaced by each other. Equations~(\ref{eq:2_6}) and (\ref{eq:2_7}) have similar forms. It will be seen that this similarity is reflected in the nonlinear viscoelasticity of the sheared supercooled liquids.

\section{SIMULATION}

Computer simulation provides an opportunity for understanding the macroscopic properties of experimental interest from the microscopic states of liquids \cite{allen1}. In this work, we performed three-dimensional BD simulations upon 20000 particles under steady Couette flow. To suppress the shear-induced crystalline ordering, a binary mixture of particles, including $N_\text{s}=4000$ small particles and $N_\text{b}=16000$ big particles, was adopted \cite{kob1}. The diameter ratio was set to be $d_\text{s}/d_\text{b}=2/3$. The equation of particle motion is expressed as \cite{brady1, allen1, ermak1}
\begin{equation}
\begin{aligned}
\bm{r}_i(t+\Delta t)-\bm{r}_i(t)=\frac{D_0}{k_\text{B}T}\bm{f}_i(t)\Delta t\\
+\sqrt{2D_0\Delta t}\bm{G}+\mathbf{H}\cdot\bm{r}_i(t)\Delta t,\label{eq:3_1}
\end{aligned}
\end{equation}
where $\bm{r}_i(t)$ is the position of particle $i$ at time $t$, $\Delta t$ is the simulation time step, $D_0$ is the Stokes-Einstein self-diffusion coefficient of particle, $k_\text{B}$ is the Boltzmann constant, $\bm{f}_i(t)$ is the deterministic non-hydrodynamic force exerted on particle $i$ caused by the interparticle potential, $\mathbf{H}$ is the strain rate tensor, and $\bm{G}$ represents the random Brownian displacement with each component an independent Gaussian variable of zero mean and unit variance. The “sliding brick” periodic boundary condition proposed by Lees and Edward is applied in accordance with the Couette flow geometry \cite{lees1}. The effective interparticle potential of charge-stabilized colloidal suspensions can be modeled by the hard-sphere Yukawa potential \cite{nagele1, nagele2}. The “potential-free” algorithm \cite{brady1, heyes1} is employed to describe the hard core in the potential. The Yukawa potential is used to represent the electrostatic interaction and is written as \cite{nagele2}
\begin{equation}
V_\text{Y}(r)=K\frac{\textrm{e}^{-z(r-d_{ij})}}{r/d_{ij}},r\ge d_{ij}\equiv(d_i+d_j)/2,\label{eq:3_2}
\end{equation}
where $d_i$ is the diameter of particle $i$, the parameters $z$ and $K$ are determined from our previous SANS data analysis \cite{wang3} and are given by $z=4.86/d_\text{b}$ and $K=9.69k_\text{B}T$, respectively. The potential function is truncated at $r=5d_\text{b}$.

Simulations at different volume fractions of particle $\phi=42.5\%, 45\%, 47.5\%$ and various shear rates were carried out. Adequate time steps were simulated to guarantee that enough data had been collected after the system approached steady state. In this section, the space and time are measured in units of $d_\text{b}$ and $\tau_0=d_\text{b}^2/D_0$, respectively. The dimensionless bare P\'eclet number $\textrm{Pe}=\dot\gamma d_\text{b}^2/4D_0$ \cite{oswald1} is sometimes used to represent the shear rate $\dot\gamma$. The long-time self-diffusion coefficient $D_\text{LS}$ at $\phi=42.5\%$ is found to be $D_\text{LS}=0.06D_0$, which is well below the dynamical criterion for freezing of colloids \cite{lowen1}, suggesting that the system is in supercooled state. 

\begin{figure}
\includegraphics[scale=1]{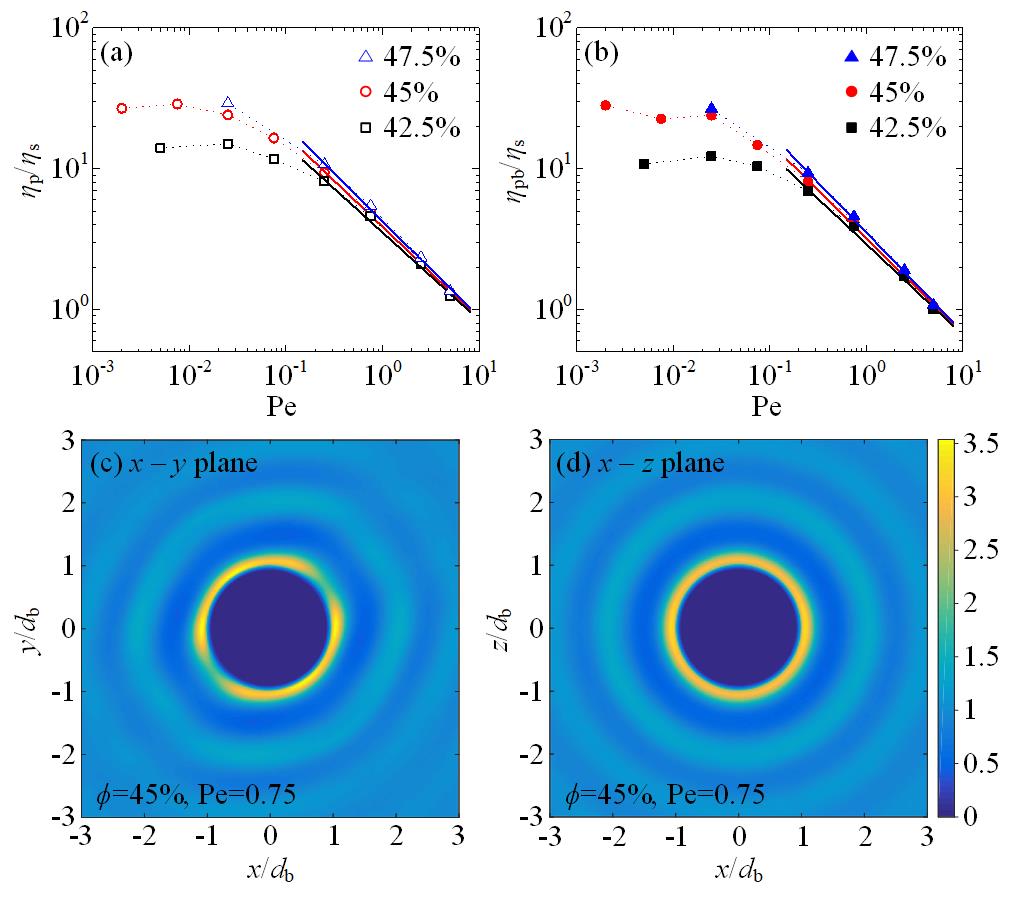}
\caption{(a) Shear viscosity contributed by the interparticle Yukawa potential ($\eta_\text{p}$). (b) Partial shear viscosity contributed by the Yukawa interactions only between big particles ($\eta_\text{pb}$). Both $\eta_\text{p}$ and $\eta_\text{pb}$ are normalized by the solvent viscosity $\eta_\text{s}$. Solid lines denote the fits with the power law $\textrm{Pe}^{-\lambda}$ in the shear thinning regime. (c) and (d) respectively display the pair distribution function of big particles, $g(\bm{r})$, at the flow - gradient ($\bm{v}-\nabla\bm{v}$ or $x-y$) plane and the flow - vorticity ($\bm{v}-\nabla\times\bm{v}$ or $x-z$) plane at the condition of $\phi=45\%$ and $\textrm{Pe}=0.75$. The thickness is $0.8d_\text{b}$ for both slices. No noticeable layering or crystallization is seen here.
\label{fig:1}}
\end{figure}

\begin{figure*}
\includegraphics[scale=1]{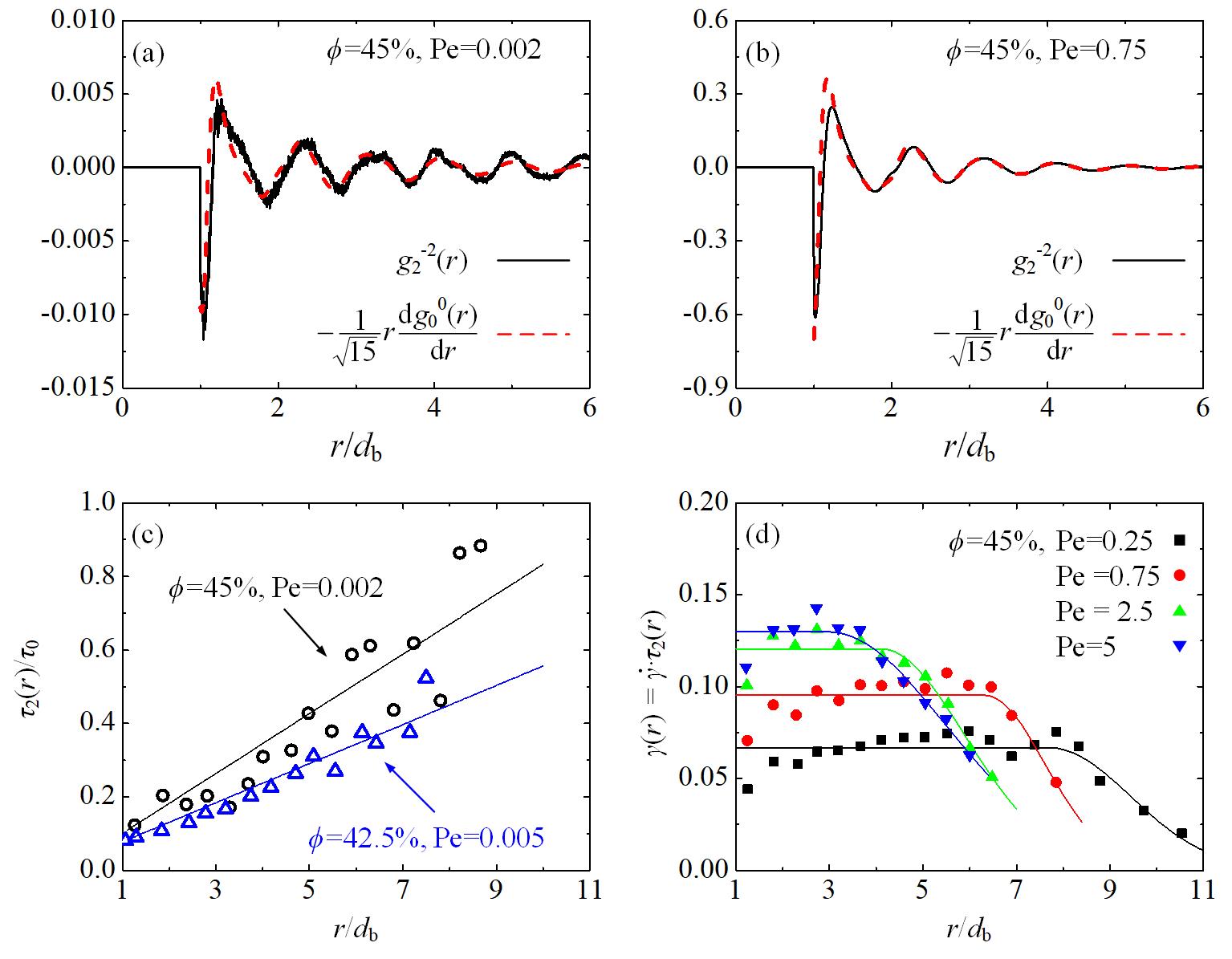}
\caption{(a) and (b) display $g_2^{-2}(r)$ as well as $-(1/\sqrt{15})r\textrm{d}g_0^0(r)/\textrm{d}r$ of the $\phi=45\%$ system at $\textrm{Pe}=0.002$ (Newtonian regime) and $0.75$ (shear thinning regime), respectively. The magnitude of $-(1/\sqrt{15})r\textrm{d}g_0^0(r)/\textrm{d}r$ is rescaled to match that of $g_2^{-2}(r)$ in both panels. (c) $\tau_2(r)$ at the conditions of $\phi=42.5\%$, $\textrm{Pe}=0.005$ and $\phi=45\%$, $\textrm{Pe}=0.002$ (Newtonian regime). Solid lines denote the linear fits. (d) $\gamma(r)=\dot\gamma\tau_2(r)$ of the $\phi=45\%$ system at $\textrm{Pe}=0.25$, $0.75$, $2.5$, and $5$ (shear thinning regime). Solid lines denote the fitted curves with Eq.~(\ref{eq:3_4}).
\label{fig:2}}
\end{figure*}

In concentrated charge-stabilized colloidal suspensions, the major source of the shear viscosity comes from the interparticle Yukawa potential \cite{morris1} and is calculated according to \cite{allen1}
\begin{equation}
\eta_\text{p}=-\frac{1}{V\dot\gamma}\bigg\langle\sum_{i=1}^Nr_{i,x}f_{i,y}\bigg\rangle,\label{eq:3_3}
\end{equation}
where $V$ is the system volume, and $f_{i,y}$ denotes the $y$-component of the Yukawa force exerted on particle $i$. The results of $\eta_\text{p}$ at different shear rates and volume fractions are shown in Fig.~\ref{fig:1}(a). The partial shear viscosity $\eta_\text{pb}$, arising from the pair Yukawa interaction between only the big particles, is also shown in Fig.~\ref{fig:1}(b). For the studied binary mixture, $\eta_\text{pb}$ contributes to more than 80\% of $\eta_\text{p}$, suggesting that the distribution and interaction of big particles play the dominant role in the nonlinear viscoelasticity of the system. Both $\eta_\text{p}$ and $\eta_\text{pb}$ smoothly transform from the Newtonian regime to shear thinning as shear rate increases. At large shear rates, the shear thinning behavior can be described by a power law $\eta\propto\dot\gamma^{-\lambda}$, as illustrated in Fig.~\ref{fig:1}(a) and (b). $\lambda$ is found to be about 0.7 for all shown volume fractions, which are slightly smaller than that of a flowing hard-sphere-like colloidal glass \cite{poon1}. The pair distribution function of big particles, denoted as $g(\bm{r})$, at $\phi=45\%$ and $\textrm{Pe}=0.75$ is displayed in Fig.~\ref{fig:1}(c) and (d). Panels c and d show $g(\bm{r})$ at the flow - gradient ($\bm{v}-\nabla\bm{v}$ or $x-y$) plane and the flow - vorticity ($\bm{v}-\nabla\times\bm{v}$ or $x-z$) plane, respectively. For all conditions given in Fig.~\ref{fig:1}, we did not observe any significant shear-induced long-range ordering at both planes. Layering effect is found to become noticeable at $\textrm{Pe}>5$. Whether this ordering is necessary for shear thinning is in debate \cite{wagner2, cohen1, hoffman1, rice1}. From our simulation, it is seen that the system is already in the $\eta\propto\dot\gamma^{-\lambda}$ regime without the appearance of significant layering. Therefore, the onset of shear thinning should be attributed to some other mechanism.

\subsection{Local elasticity}

The remarkable shear-induced anisotropy shown in the $x-y$ plane (Fig.~\ref{fig:1}(c)) is mainly due to the nonzero $g_2^{-2}(r)$. Figure~\ref{fig:2}(a) and (b) show the $g_2^{-2}(r)$ of the $\phi=45\%$ system at $\textrm{Pe}=0.002$ and $0.75$, respectively in the Newtonian regime and the shear thinning regime. The profiles of $-(1/\sqrt{15})r\textrm{d}g_0^0(r)/\textrm{d}r$ at same conditions are also shown. According to Eq.~(\ref{eq:2_6}), $\tau_2(r)$ connects these two functions and thus is crucial for characterizing the microstructural distortion. As shown in Fig.~\ref{fig:2}(a) and (b), the characteristic variations of these two functions are generally in phase. Therefore, we can depict the profile of $\tau_2(r)$ by simply dividing $g_2^{-2}(r)$ by $-(1/\sqrt{15})r\textrm{d}g_0^0(r)/\textrm{d}r$ at each peak position. Some results of $\tau_2(r)$ are given in Fig.~\ref{fig:2}(c) and (d). Panel (c) shows the $\tau_2(r)$ at $\phi=45\%$, $\textrm{Pe}=0.002$ and $\phi=42.5\%$, $\textrm{Pe}=0.005$, both are in the Newtonian regime. While panel (d) shows the profiles of $\dot\gamma\tau_2(r)$ of the $\phi=45\%$ system at $\textrm{Pe}=0.25$, $0.75$, $2.5$, and $5$, which are all in the shear thinning regime described by $\eta\propto\dot\gamma^{-\lambda}$.

Though our emphasis is the nonlinear rheology, it is inspiring to have a glance at the Newtonian regime first. Seen from Fig.~\ref{fig:2}(c), $\tau_2(r)$ at $\phi=42.5\%$, $\textrm{Pe}=0.005$ depends on $r$ linearly. The profile of $\tau_2(r)$ at $\phi=45\%$, $\textrm{Pe}=0.002$ also exhibits an increasing trend as $r$ increases, which can be roughly described by a linear relation. We fit these two $\tau_2(r)$ by $\tau_2(r)=\tau_\text{s}(r/r_1)+\tau_\text{c}$, where $r_1$ is the position of the first positive peak of $g_2^{-2}(r)$, $\tau_\text{s}$ and $\tau_\text{c}$ are fitting parameters. Since $r_1$ is close to $d_\text{b}$, $\tau_2(r_1)=\tau_\text{s}+\tau_\text{c}$ gives the characteristic time for the relaxation of the anisotropy of the cage. For both of these two cases, $\tau_2(r_1)$ is close to the Maxwell relaxation time $\tau_\text{M}$. The linear behavior of $\tau_2(r)$ in the Newtonian regime is similar to the observation in a simulation study of equilibrium atomic liquids \cite{egami4}. In that work, the authors found that the relaxation time $\tau_\text{v}(r)$ of the van Hove correlation function increases linearly with distance $r$ \cite{egami4}. It can be understood by the argument that $\tau_\text{v}(r)$ should scale with the thermal-activated particle number fluctuation $\Delta N(r)$, which behaves as $\Delta N(r)\propto r\sqrt{4\pi\rho g(r)}$ according to the central limit theorem \cite{egami4}. At large $r$, $g(r)\to1$ and consequently $\tau_\text{v}(r)\propto\Delta N(r)\propto r$. Such analogy suggests that the shear-induced microscopic anisotropy in the Newtonian regime is relaxed by the thermal fluctuation of particles, which is as expected.

\begin{figure}
\includegraphics[scale=1]{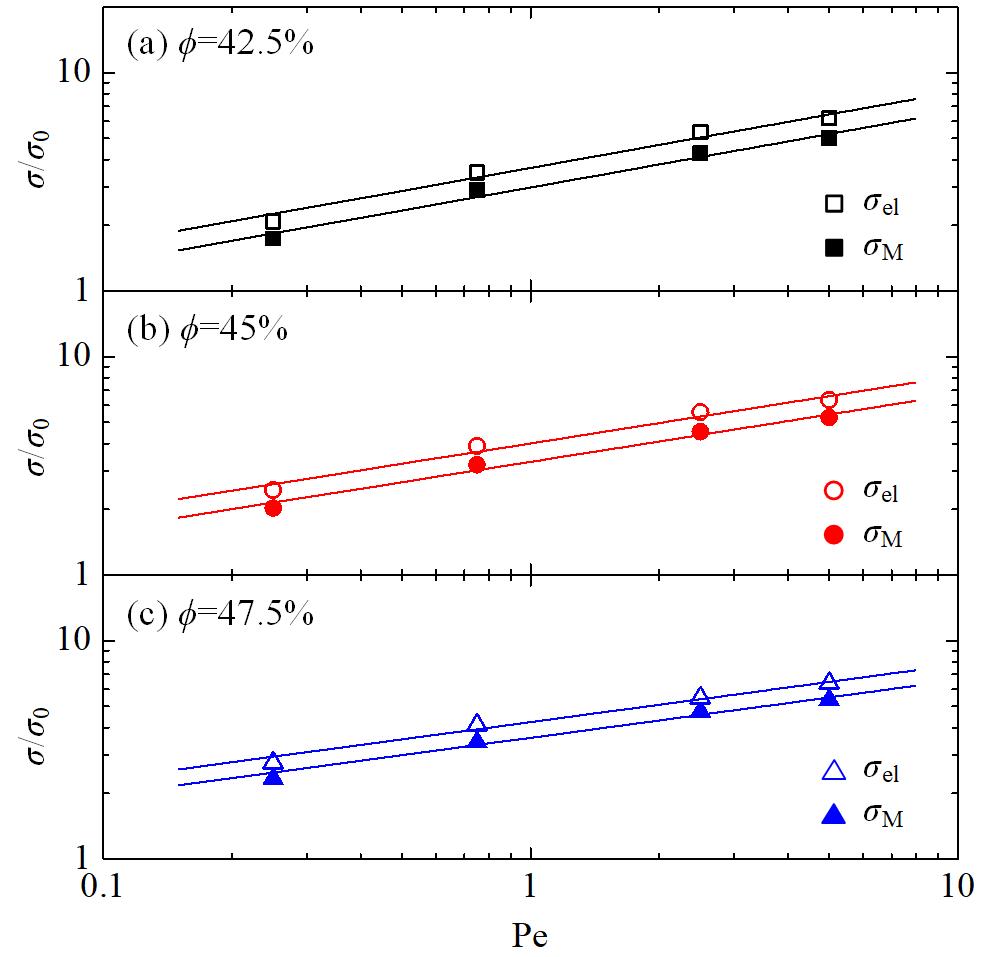}
\caption{The shear stress $\sigma_\text{M}=\dot\gamma\eta_\text{pb}$ (solid symbols) and the microscopic elastic stress $\sigma_\text{el}=G_\infty\gamma_\text{NN}$ (open symbols) in the shear thinning regime. Lines denote the fits with the power law $\sigma\propto\textrm{Pe}^\epsilon$. The stress is in unit of $\sigma_0=4k_\text{B}T/3\pi d_\text{b}^3$. With this unit, the stress can be expressed as $\sigma/\sigma_0=\textrm{Pe}\cdot(\eta/\eta_\text{s})$.
\label{fig:3}}
\end{figure}

Figure~\ref{fig:2}(d) shows the profiles of $\dot\gamma\tau_2(r)$ in the shear thinning regime. In contrast to the cases in the Newtonian regime, here $\tau_2(r)$ exhibits a plateau spanning several $d_\text{b}$. We denote the range of this plateau by $\xi$. For $r\lesssim\xi$, $\dot\gamma\tau_2(r)$ can be approximated by a constant $\bar\gamma$, and Eq.~(\ref{eq:2_6}) reduces to a form akin to Eq.~(\ref{eq:2_7}) that describes the anisotropy induced by small elastic deformation. To extract $\xi$ and $\bar\gamma$, we fit $\dot\gamma\tau_2(r)$ with the following equation
\begin{equation}
\gamma(r)=\dot\gamma\tau_2(r)=
\left\{
\begin{aligned}
&\bar\gamma, &r\le g\\
&\bar\gamma\exp{\Big[-\frac{(r-g)^2}{2a^2}\Big]}, &r>g
\end{aligned}
\right. ,\label{eq:3_4}
\end{equation}
where $\bar\gamma$, $g$, and $a$ are fitting parameters. $\xi$ is then obtained by $\xi=g+\sqrt{2\ln 2}a$. As the shear rate increases, the plateau value $\bar\gamma$ enhances while its spatial range $\xi$ shrinks. Such plateau has been observed in a simulation study of sheared metallic glasses \cite{egami2}. Because of its similarity to Eq.~(\ref{eq:2_7}), the authors of Ref. \cite{egami2} identified the plateau as a region of elastic response \cite{egami2}: within the spatial range of this region, the local structure undergoes an elastic deformation with an average strain given by $\bar\gamma$ when the system is under steady shear. This local solid-like response survives only for a lifetime about $2\bar\gamma/\dot\gamma$, and then is relaxed by flow. We call this region the localized elastic region (LER)\footnote{In our previous preliminary Rheo-SANS analysis, we named this localized region of elastic response as the “transient elasticity zone” (see Ref. \cite{wang3}). We feel that “localized elastic region” is more proper.}. In this picture, the major source of shear stress comes from the elastic deformation of the LER:
\begin{equation}
\dot\gamma\eta_\text{pb}\approx G_\infty\gamma_\text{NN},\label{eq:3_5}
\end{equation}
where $\gamma_\text{NN}$ is the strain of the nearest neighbors ($\gamma_\text{NN}$ equals to the value of $\gamma (r)$ at the first shell), and $G_\infty$ is the infinite shear modulus. In the preceding equation, we choose $\gamma_\text{NN}$ rather than $\bar\gamma$ to calculate the particle-level stress of the reference particle \cite{egami5}, because $\gamma_\text{NN}$ gives a better description on the local strain around the reference particle. As for $G_\infty$, in the shear flow, it can be approximated by the angle-averaged modulus given by \cite{hoover1}
\begin{equation}
G_\infty=\frac{2\pi}{15}\rho_\text{b}^2\int{\big[4r^3V_\text{Y}'(r)+r^4V_\text{Y}''(r)\big]g_0^0(r)}\textrm{d}r,\label{eq:3_6}
\end{equation}
where $\rho_\text{b}$ is the number density of big particles. Figure~\ref{fig:3} examines the validity of Eq.~(\ref{eq:3_5}) by comparing the shear stress $\sigma_\text{M}=\dot\gamma\eta_\text{pb}$ contributed by interparticle potential and the microscopic elastic stress $\sigma_\text{el}=G_\infty\gamma_\text{NN}$. For all points shown in Fig.~\ref{fig:3}, $\sigma_\text{M}$ equals to about 85\% of $\sigma_\text{el}$. We fit both $\sigma_\text{M}(\textrm{Pe})$ and $\sigma_\text{el}(\textrm{Pe})$ with the power law $\sigma\propto\textrm{Pe}^\epsilon$, and find that the exponents ($\epsilon$) for $\sigma_\text{M}(\textrm{Pe})$ and for $\sigma_\text{el}(\textrm{Pe})$ well agree with each other. The resemblance between $\sigma_\text{M}$ and $\sigma_\text{el}$ supports the idea that the localized elasticity governs the rheology in the shear thinning regime. The difference between $\sigma_\text{M}$ and $\sigma_\text{el}$ shows that such simple elastic model overestimates the stress, and should be attributed to the yielding and rearrangement of the local structure induced by the nonaffine displacement of particles \cite{hents1, wittmer1, zaccone1}. Note that, $G_\infty$ is not sensitive to the shear rate in our simulation. Suggested by Eq.~(\ref{eq:3_5}) and the results given in Fig.~\ref{fig:3}, $\gamma_\text{NN}$ is expected to depend on the shear rate by $\gamma_\text{NN}\propto\dot\gamma^\epsilon$ with $\epsilon\approx 1-\lambda$.

\begin{figure}
\includegraphics[scale=1]{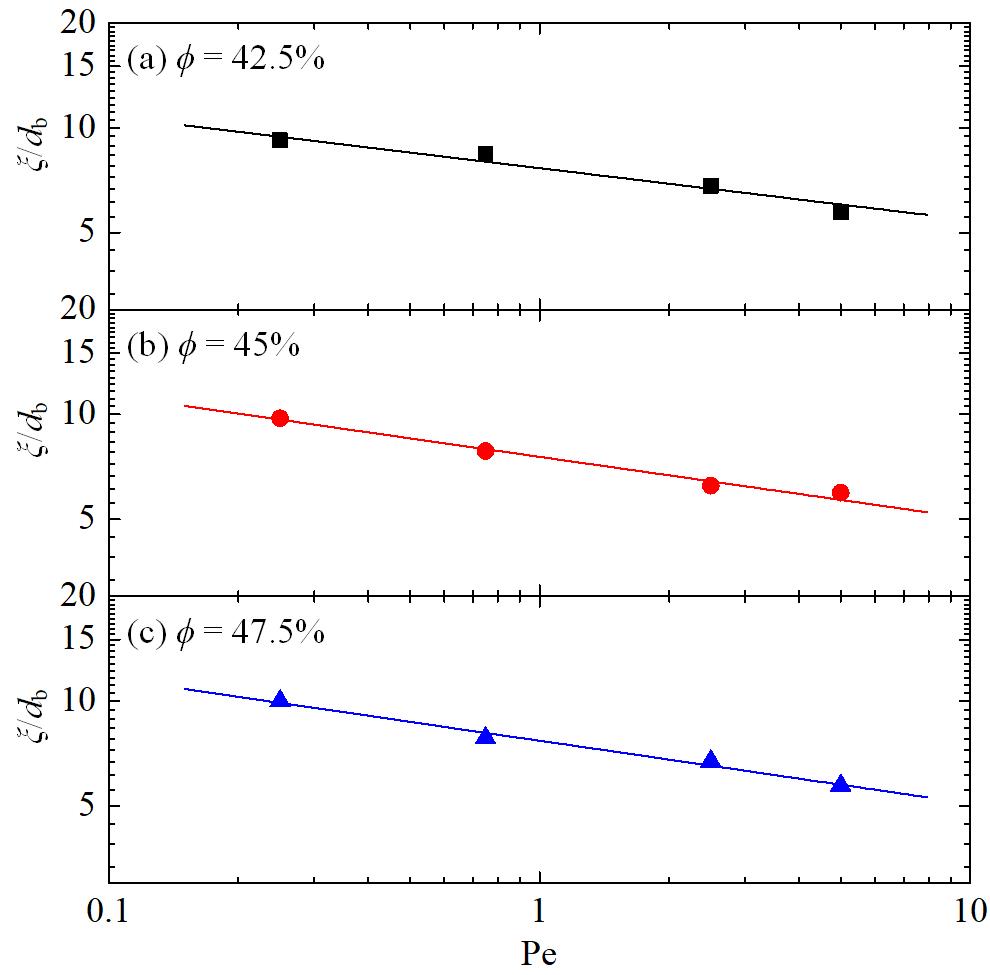}
\caption{$\xi(\textrm{Pe})$ in the shear thinning regime. Symbols denote the simulated results. Lines denote the fits with the power law $\xi\propto\textrm{Pe}^{-\nu}$.
\label{fig:4}}
\end{figure}

Despite the good agreement shown in Fig.~\ref{fig:3}, the existence and properties of the LER in sheared supercooled liquids call for further investigations. Particularly, the spatial range of the elastic response, $\xi$, should be quantitatively related to the elastic properties of the sheared liquid. In a series of papers, Dyre proposed the picture of “solidity” to explain the viscous behavior of supercooled liquids \cite{dyre1, dyre2, dyre3, dyre4}. This scenario is based on the fact that in viscous liquids, most molecular motion is purely vibrational, and the flow events are relatively rare. Therefore, between two successive flow events involving the same molecule, the local dynamics is solid-like. This “solidity” only happens within the “solidity length” $l_\text{solid}$, which can be evaluated as follows \cite{dyre2}. Set $l_0$ the characteristic length scale of a flow event. Within the range of $l_\text{solid}$, the number of possible locations for flow events is about $N_\text{f}\approx(l_\text{solid}/l_0)^3$. Denoting $\tau$ as the local relaxation time, the average time between two flow events within the solidity range is estimated by $\tau/N_\text{f}=\tau(l_0/l_\text{solid})^3$. $\tau/N_\text{f}$ should be equal to $l_\text{solid}/c$ to keep the solidity, where $c$ is the sound speed. Then we have $l_\text{solid}^4=l_0^3\tau c$. Notice that, such spatially-extended solidity does not exist in liquids at high temperatures, in which the phonons are found to be highly localized by computer simulations \cite{egami6}. We directly generalize this relation to the nonlinear regime of the sheared supercooled liquids. Considering that the relaxation of local configuration is mainly induced by external shear in the nonlinear regime, the length scale of the LER $\xi$ could be estimated as
\begin{equation}
\xi\approx l_0^{3/4}\Big(\frac{2\gamma_\text{NN}}{\dot\gamma}\Big)^{1/4}c_\text{T}^{1/4},\label{eq:3_7}
\end{equation}
where $c_\text{T}$ is the transverse sound speed. Assuming that $l_0$ and $c_\text{T}$ are not sensitive to $\dot\gamma$, Eq.~(\ref{eq:3_7}) results in a power law of $\xi\propto\dot\gamma^{-\nu}$ with $\nu=\lambda/4$. Figure~\ref{fig:4} displays the power-law fit of $\xi(\dot\gamma)$ in the shear thinning regime.

Above analysis shows that the three exponents in the shear thinning regime, $\lambda$ in $\eta\propto\dot\gamma^{-\lambda}$, $\epsilon$ in $\gamma_\text{NN}\propto\dot\gamma^\epsilon$, and $\nu$ in $\xi\propto\dot\gamma^{-\nu}$, are related by
\begin{equation}
\lambda=4\nu=1-\epsilon.\label{eq:3_8}
\end{equation}
Note that, $\lambda$ describes the macroscopic feature of shear thinning, $\epsilon$ describes the amplitude of the shear-induced microstructural distortion, and $\nu$ describes the spatial range of the elastic response. Therefore, this relation connects the bulk nonlinear rheology, the microscopic structure, and the elastic properties of the system. Table 1 lists the values of $\lambda$, $\epsilon$ and $\nu$ of all simulated volume fractions. It is seen that Eq.~(\ref{eq:3_8}) works very well.

\begin{table}[h]
\caption{\label{tab:table1}
The values of $\lambda$, $\epsilon$ and $\nu$
}
\begin{ruledtabular}
\begin{tabular}{cccccc}
$\phi$ & $\nu$ & $\epsilon$ & $\lambda$ & $4\nu$ & $1-\epsilon$ \\
\colrule
47.5\% & 0.178 & 0.272 & 0.709 & 0.713 & 0.728 \\
45\%   & 0.170 & 0.329 & 0.680 & 0.681 & 0.671 \\
42.5\% & 0.158 & 0.356 & 0.645 & 0.632 & 0.644 \\
\end{tabular}
\end{ruledtabular}
\end{table}

A fundamental difference between the Newtonian regime and the shear thinning regime lies in the way of local structural rearrangement. In the shear thinning regime, the rearrangement of local configuration is driven by the imposed shear. While in the Newtonian regime, the shear rate is too slow to compete with the spontaneous relaxation. In this case, the local structural relaxation is thermal-activated, and the relaxation time linearly depends on $r$. Thus, though the “solidity” exists, one cannot identify an extended-range zone within which the particles move in a highly coherent way.

\begin{figure}[!htb]
\includegraphics[scale=1]{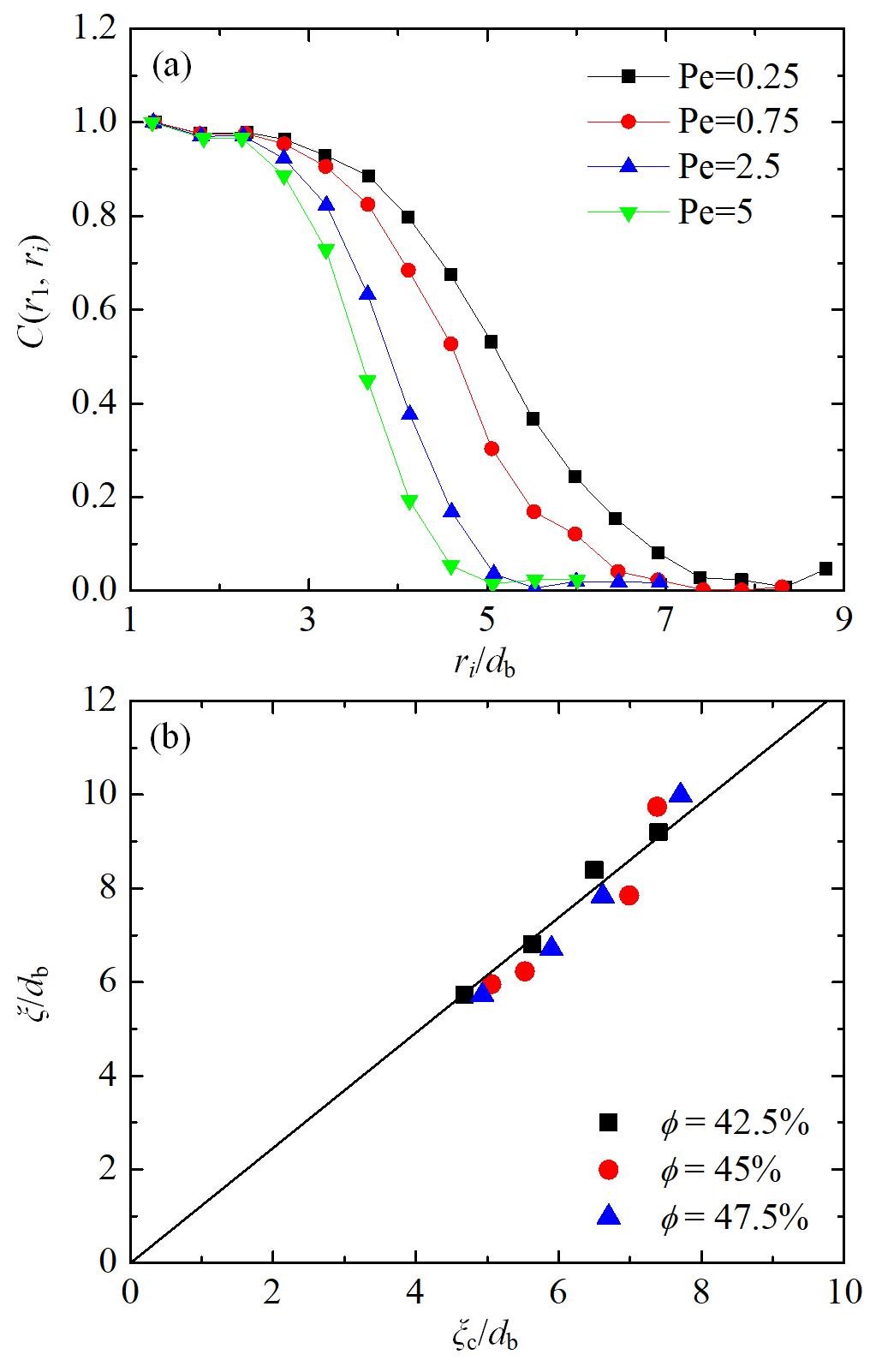}
\caption{(a) $C(r_1,r_i)$ as a function of $r_i$ for the $\phi=45\%$ system at $\textrm{Pe}=0.25$, $0.75$, $2.5$ and $5$. In this case, $C(r_1,r_i)$ becomes statistically noisy at $C(r_1,r_i)\lesssim2\%$. Therefore, we set 2\% as the “effective zero” for $C(r_1,r_i)$. (b) Scatterplot of $\xi$ and $\xi_\text{c}$ for all simulated points in shear thinning regime. The solid line denotes the relation of $\xi=1.23\xi_\text{c}$.
\label{fig:5}}
\end{figure}

According to the above picture, the particle displacement exhibits elastic coherency within $r\lesssim\xi$, and becomes uncorrelated at $r\gtrsim\xi$. We can test this statement with the correlation between the transient intensities of different peaks of $g_2^{-2}(r)$. To perform this test, we define the transient pair distribution function for particle $j$ at time $t$ by:
\begin{equation}
g(\bm{r},j,t)=\frac{1}{\rho}\sum_{i\ne j}\delta\big(\bm{r}-[\bm{r}_i(t)-\bm{r}_j(t)]\big).\label{eq:3_9}
\end{equation}
Extracting its SHE coefficient with $l=2$ and $m=-2$, we obtain $g_2^{-2}(r_i;j,t)$, the intensity of the $i$th peak of the transient $g_2^{-2}(r)$ of particle $j$ at time $t$ ($r_i$ denotes the position of the $i$th peak of $g_2^{-2}(r)$). To enhance the statistics, we divide all 16000 big particles into 50 groups according to the ascending sequence of the value of $g_2^{-2}(r_1;j,t)$, and calculate the average value of $g_2^{-2}(r_i;j,t)$ for each group, which is written as $\langle g_2^{-2}(r_i,t)\rangle_k$ for the $k$th group. Then, we can evaluate the correlation between the transient intensity of the first positive peak of $g_2^{-2}(r)$ and that of the peak at $r=r_i$ with the following function:
\begin{widetext}
\begin{equation}
C(r_1,r_i)=\Bigg|\Big\langle\frac
{
E\lbrace[\langle g_2^{-2}(r_1,t)\rangle_k-g_2^{-2}(r_1)][\langle g_2^{-2}(r_i,t)\rangle_k-g_2^{-2}(r_i)]\rbrace
}
{
\sqrt{E[\langle g_2^{-2}(r_1,t)\rangle_k^2]-[g_2^{-2}(r_1)]^2}
\sqrt{E[\langle g_2^{-2}(r_i,t)\rangle_k^2]-[g_2^{-2}(r_i)]^2}
}
\Big\rangle_t\Bigg|
,\label{eq:3_10}
\end{equation}
\end{widetext}
where $E(A)$ denotes the average of $A$ over all groups and is given by $E(A)=\sum_{k=1}^{50}A_k/50$, and $\langle\dots\rangle_t$ means the average about $t$. Figure~\ref{fig:5}(a) displays $C(r_1,r_i)$ for the $\phi=45\%$ system at $\textrm{Pe}=0.25$, $0.75$, $2.5$ and $5$. The behaviors of $C(r_1,r_i)$ can be summarized as follows: (i) $C(r_1,r_i)$ steadily decreases as $r_i$ increases, and becomes negligible as $r_i$ approaches $\xi$. This profile clearly reveals the deformation heterogeneity within the LER. (ii) Similar to $\xi$, the range of coherency shrinks with shear rate. We define the coherency length $\xi_\text{c}$ of $C(r_1,r)$ by the value of  $r$ at which $C(r_1,r)$ decays to effective zero, and compare $\xi_\text{c}$ with $\xi$ in Fig.~\ref{fig:5}(b). Linearly fitting all shown points leads to a relation $\xi=1.23\xi_\text{c}$. The Pearson correlation coefficient \cite{pearson} between $\xi_\text{c}$ and $\xi$ is 0.98, indicating a good linear correlation. The agreement between $\xi_\text{c}$ and $\xi$ confirms the localization of the elastic response. The fact that $\xi_\text{c}$ is slightly smaller than $\xi$ can be understood by noticing that $\gamma(r)$ starts deviating from $\bar\gamma$ at $r<\xi$.

Summarizing above results, we give the following microscopic mechanism for the nonlinear rheology of sheared supercooled liquids: In response to the imposed shear, the particles within a limited spatial range, which we term as the LER, undergo concerted elastic deformation with a certain lifetime. LER is the structural unit that resists the external shear, as suggested by the agreement between the shear stress and the average elastic stress sustained by LER. The elastic coherency of particle displacement decreases as the distance from the reference particle increases, and disappears at the periphery of the LER. The deformation and yielding of LER are ubiquitous and persistently successive in the flow. Note that, it is important to identify the mesoscopic structural unit that store and release the elastic energy in viscoelastic materials \cite{cates1}. Thus, LER is conceptually valuable for understanding the nonlinear viscoelasticity of supercooled liquids.

\subsection{Local structural rearrangement}

\begin{figure}
\includegraphics[scale=1]{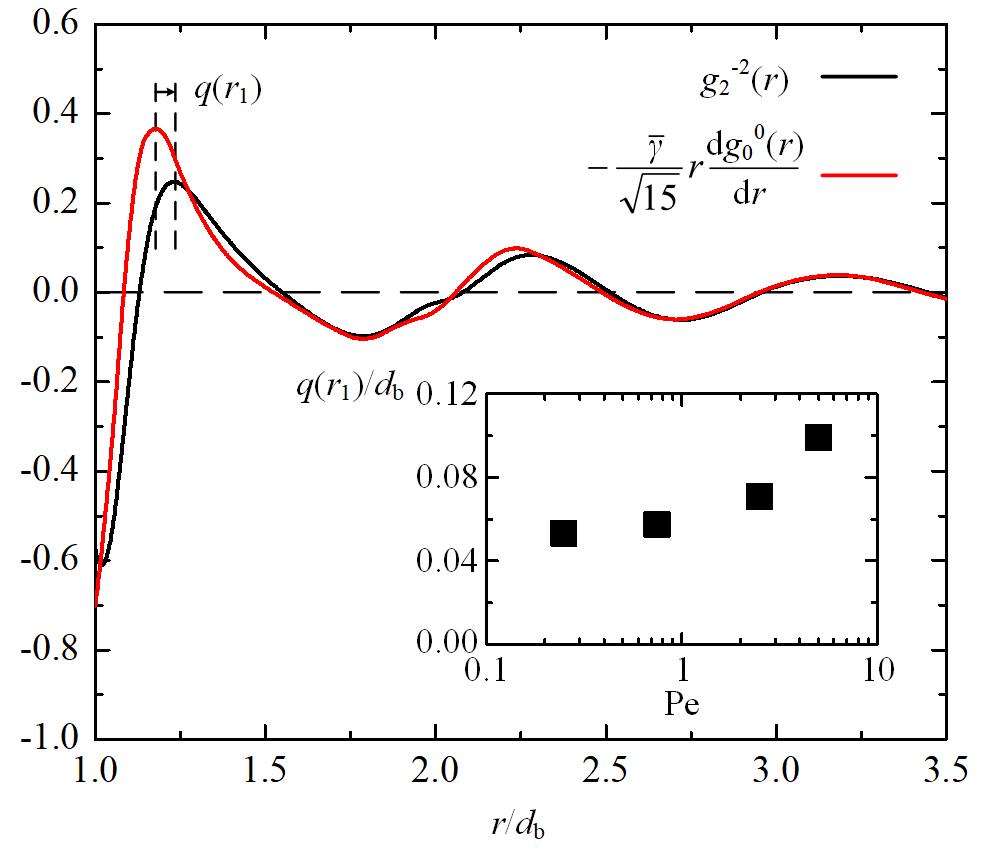}
\caption{Comparison between $-(\bar\gamma/\sqrt{15}) r\textrm{d}g_0^0(r)/\textrm{d}r$ and $g_2^{-2}(r)$ at $\textrm{Pe}=0.75$ and $\phi=45\%$. The phase difference $q(r_1)$ at the first peak is denoted. Inset: $q(r_1)$ as a function of $\textrm{Pe}$ in the shear thinning regime of the $\phi=45\%$ system.
\label{fig:6}}
\end{figure}

In sub-section III.A, we have established the existence of the LER in the shear thinning regime. In this sub-section, we are going to explore the relaxation of the LER, which is another fundamental aspect of the local viscoelasticity. We start our discussion by comparing $g_2^{-2}(r)$ and $-(\bar\gamma/\sqrt{15}) r\textrm{d}g_0^0(r)/\textrm{d}r$ in detail. According to the analysis in sub-section III.A, these two functions are similar to each other to some extent, which leads to the identification of the LER. However, in principle they should be different, because $g_2^{-2}(r)$ and $-(\bar\gamma/\sqrt{15}) r\textrm{d}g_0^0(r)/\textrm{d}r$ respectively represent the microscopic distortion of a liquid and that of an elastic solid. An important distinction is that the characteristic variations of these two functions exhibit a phase difference. An example is given in Fig.~\ref{fig:6}, where we denote the phase difference at the first positive peak of $g_2^{-2}(r)$ as $q(r_1)$. $q(r_1)$ enhances as the shear rate increases, as shown in the inset of Fig.~\ref{fig:6}. In addition, $q(r)$ exhibits a descending trend as $r$ increases. The appearance of $q(r)$ signifies the nonaffine particle motions.

\begin{figure}
\includegraphics[scale=1]{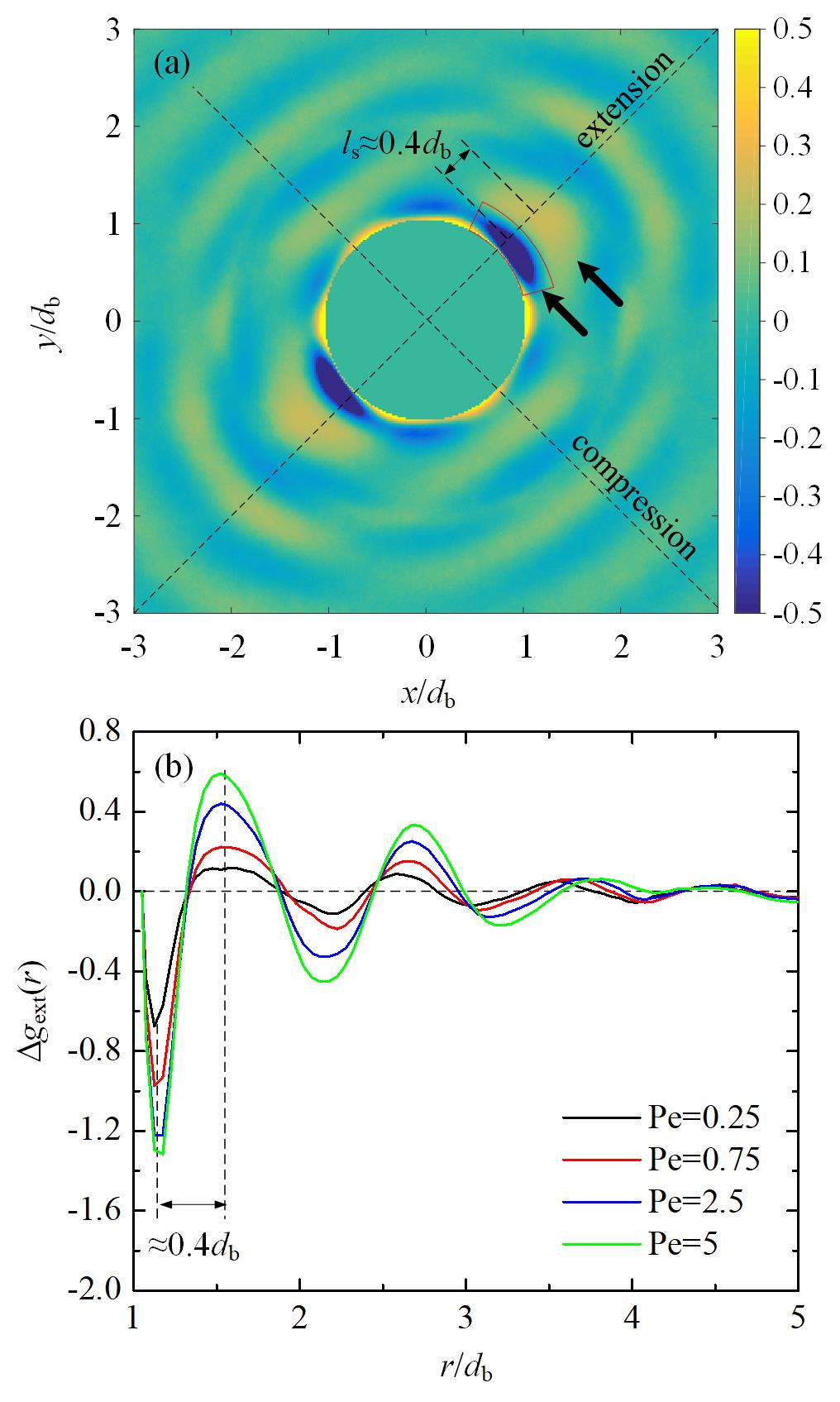}
\caption{(a) The slice of $\Delta g(\bm{r})=g_\text{liq}(\bm{r})-g_\text{aff}(\bm{r})$ in the $x-y$ plane. $g_\text{liq}(\bm{r})$ is given by the pair distribution function of the flow at $\textrm{Pe}=0.75$ and $\phi=45\%$. $g_\text{aff}(\bm{r})$ is the pair distribution function obtained by affinely shearing the equilibrium structure of the $\phi=45\%$ system with the strain of $\bar\gamma=0.095$. The thickness of the slice is $0.8d_\text{b}$. Two arrows denote the basin and the peak on the extensional axis. (b) The profiles of $\Delta g(\bm{r})$ along the extensional axis, $\Delta g_\text{ext}(r)$, at $\textrm{Pe}=0.25$, $0.75$, $2.5$ and $5$ for the $\phi=45\%$ system. The first negative peak and the first positive peak of $\Delta g_\text{ext}(r)$ respectively corresponds to the basin and peak denoted by arrows in panel (a). The distance between these two peaks is about $0.4d_\text{b}$.
\label{fig:7}}
\end{figure}

To visualize the yielding process of LER, we plot the difference (($\Delta g(\bm{r})$) between the pair distribution function of a sheared liquid ($g_\text{liq}(\bm{r})$) and that of an affinely deformed system ($g_\text{aff}(\bm{r})$) in the $x-y$ plane in Fig.~\ref{fig:7}(a) ($\Delta g(\bm{r})=g_\text{liq}(\bm{r})-g_\text{aff}(\bm{r})$). The simulated $g(\bm{r})$ of the $\phi=45\%$ system at $\textrm{Pe}=0.75$ is adopted as $g_\text{liq}(\bm{r})$. $g_\text{aff}(\bm{r})$ is obtained by affinely shearing the equilibrium structure of the $\phi=45\%$ system with the strain of $0.095$, which is just the average strain $\bar\gamma$ of the LER at $\textrm{Pe}=0.75$ and $\phi=45\%$ \footnote{$g_\text{aff}(\bm{r})$ can be calculated by $g_\text{aff}(\bm{r})=g_\text{eq}(\bm{F}\cdot\bm{r})$, where the tensor $\bm{F}$ is given by $\bm{F}^{-1}=\left(\begin{array}{ccc}1&\bar\gamma&0\\0&1&0\\0&0&1\end{array}\right)$}. Similar to $q(r)$, the nonzero $\Delta g(\bm{r})$ reflects the deviation from the purely elastic deformation caused by the nonaffine particle displacements. The two-dimensional (2D) pattern of $\Delta g(\bm{r})$ in the $x-y$ plane is shown in Fig.~\ref{fig:7}(a). It exhibits variations at all azimuthal directions. The most prominent variation is along the extensional axis, where there is a basin at the first coordination shell (blue color, marked by an arrow) followed by a peak located between the first and second coordination shells (yellow color, marked by an arrow). Notice that, on the extensional axis, $\Delta g(\bm{r})$ has the same profile with $g_2^{-2}(r)-[-(\bar\gamma/\sqrt{15}) r\textrm{d}g_0^0(r)/\textrm{d}r]$ up to the first order of $\dot\gamma\tau_2$ (their difference comes from the higher-order terms, such as $g_2^0$, $g_2^{-2}$ and $g_4^4$). Therefore, the existence of these two lobes corresponds to the nonzero phase difference $q(r_1)$. The profile of $\Delta g(\bm{r})$ along the extensional axis, denoted as $\Delta g_\text{ext}(r)$, is plotted in Fig.~\ref{fig:7}(b) for the $\phi=45\%$ system. Similar to $q(r)$, the magnitude of $\Delta g_\text{ext}(r)$ increases as the shear rate increases in the shear thinning regime, and decreases as $r$ increases.

\begin{figure}
\includegraphics[scale=1]{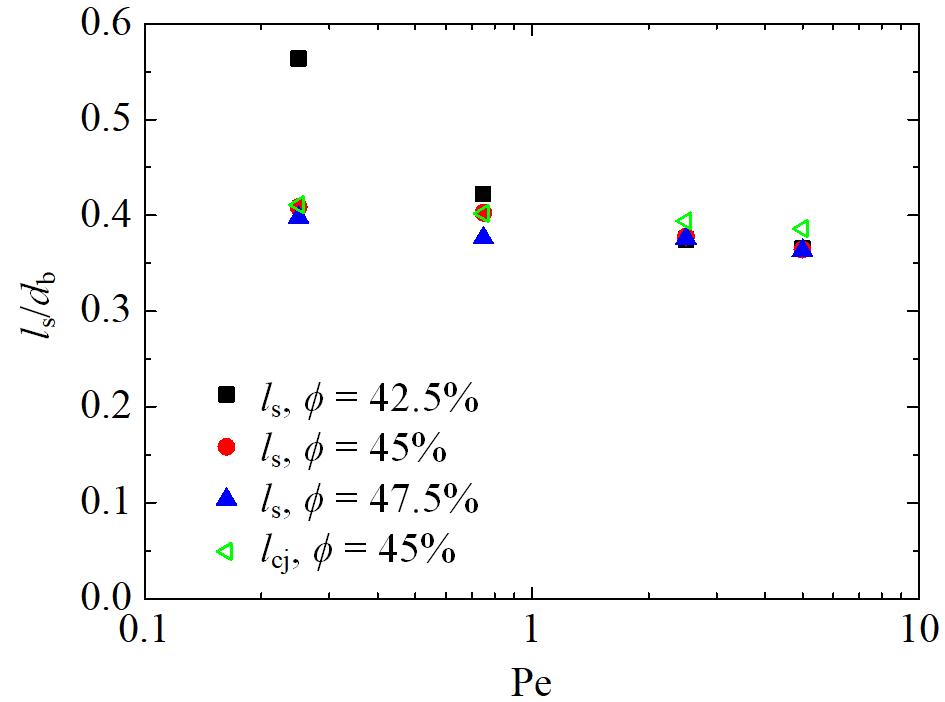}
\caption{The step lengths $l_\text{s}$ found from the slice of $\Delta g(\bm{r})$ in the $x-y$ plane for all simulated conditions in the shear thinning regime (solid symbols). The cage jump lengths $l_\text{cj}$ for the $\phi=45\%$ system (open symbols) are also shown for comparison.
\label{fig:8}}
\end{figure}

In many previous studies of quiescent atomic \cite{biroli1, biroli2} or colloidal \cite{weeks3} supercooled liquids, the cage rearrangement is attributed to the particle motions with irreversible large displacements. In Refs. \cite{biroli1, biroli2}, Candelier \textit{et al}. call such particle motion the “cage jump”. Similar mechanism has also been proposed in flowing colloids \cite{poon1, schall2}. The emergence of the basins and peaks in $\Delta g(\bm{r})$ uncovers the nonaffine particle displacement in the flow. The prominent oscillation of $\Delta g(\bm{r})$ along the extensional axis suggests that the distortion of the nearest shell is mainly relaxed by the large-step nonaffine displacement from the basin to the peak denoted by arrows in Fig.~\ref{fig:7}(a). The step length, $l_\text{s}$, is given by the distance between these two lobes. Seen from Fig.~\ref{fig:7}, $l_\text{s}$ is about $0.4d_\text{b}$. Figure~\ref{fig:8} shows the values of $l_\text{s}$ for all simulated points in the shear thinning regime. $l_\text{s}$ exhibits a decreasing dependence on shear rate, because LER is more sheared at higher shear rates, which shortens the escape path of the particle. In addition, $l_\text{s}$ decreases as the volume fraction increases, corresponding to the reduction of the free space. To justify the $l_\text{s}$ obtained from $\Delta g(\bm{r})$, we calculate the length of the “cage jump”, $l_\text{cj}$, by generalizing the method proposed by Candelier \textit{et al}. \cite{biroli1, biroli2} into the flowing state. The detail in finding $l_\text{cj}$ is given in Appendix A. The results are compared with $l_\text{s}$ in Fig.~\ref{fig:8} for the $\phi=45\%$ system. It is seen that $l_\text{s}$ and $l_\text{cj}$, found from independent methods, highly agree with each other. As the distance from the reference particle $r$ increases, such nonaffine jump becomes less significant, as indicated by the fact that both $\Delta g_\text{ext}(r)$ and $q(r)$ decrease as $r$ increases. Note that, the oscillation of $g_0^0(r)$ also decreases with $r$. The weak radial ordering at large $r$ mitigates the restriction on the movement of particle. Consequently, the relaxation of the shell can be realized more flexibly and become less direction-dependent.

\begin{figure}
\includegraphics[scale=1]{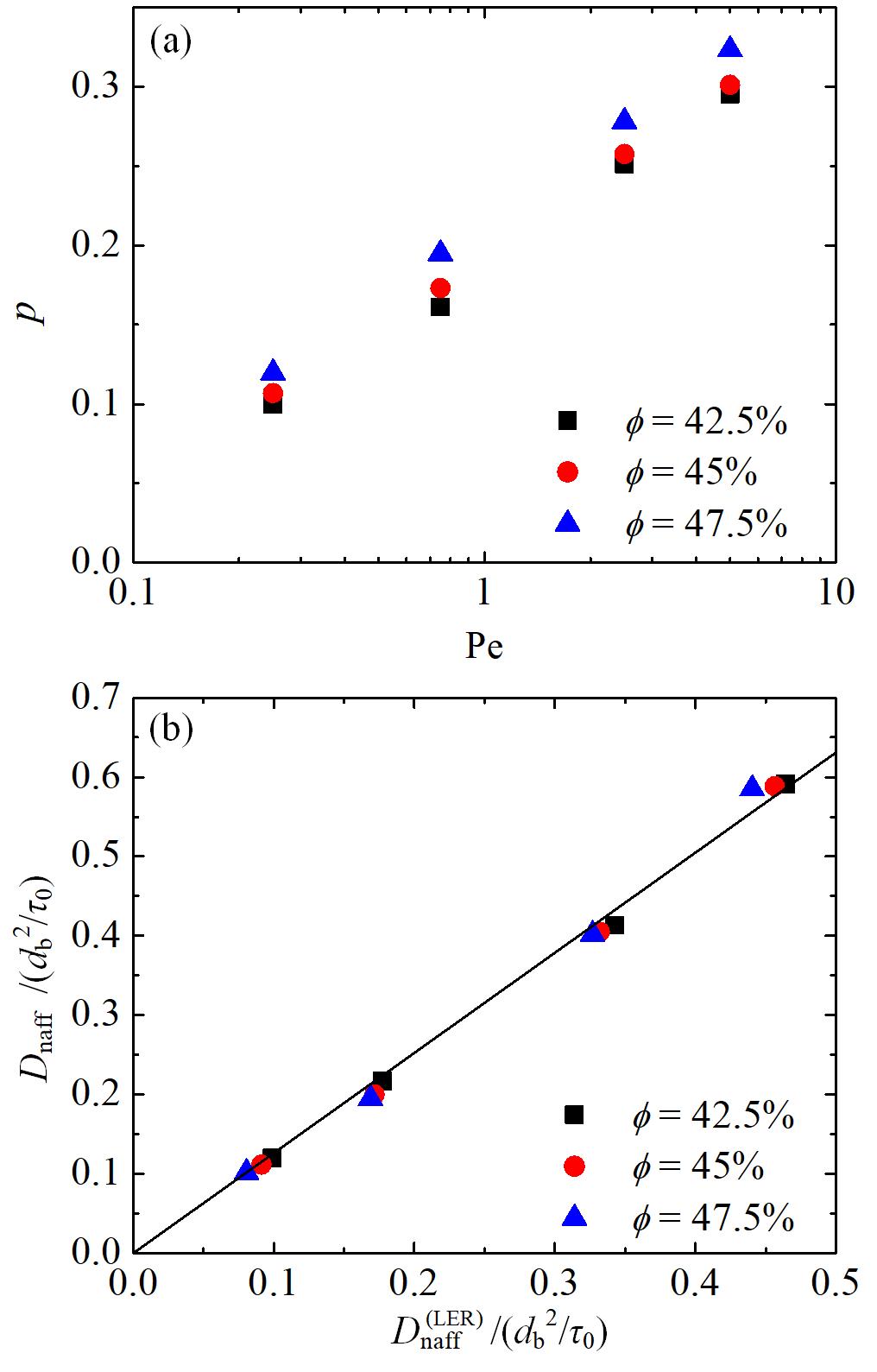}
\caption{(a) $p$ as a function of $\textrm{Pe}$ in the nonlinear regime. (b) Scatterplot of $D_\text{naff}$ and $D_\text{naff}^{\text{(LER)}}$. The solid line denotes the relation $D_\text{naff}=1.26D_\text{naff}^{\text{(LER)}}$.
\label{fig:9}}
\end{figure}

As shown in Fig.~\ref{fig:6} and Fig.~\ref{fig:7}(b), both $q(r)$ and $\Delta g_\text{ext}(r)$ grow as the shear rate increases, suggesting that the nonaffine particle motion enhances. To quantify the strength of the nonaffine motion, we calculate the following quantity for the basin on the extensional axis
\begin{equation}
p=\frac{N_\text{naff}}{N_\text{aff}}
,\label{eq:3_11}
\end{equation}
where $N_\text{naff}$ and $N_\text{aff}$ denote the particle number in certain regions. $N_\text{naff}$ is the number of big particles which leave the region of the basin through large-step nonaffine displacements. To calculate $N_\text{naff}$, we pick a region that encompasses the basin (marked by red lines in Fig.~\ref{fig:7}(a)). The volume of this region is denoted as $v_\text{b}$. Then, $N_\text{naff}$ is obtained by $N_\text{naff}=|\rho\int_{v_\text{b}}\Delta g(\bm{r})\textrm{d}\bm{r}|$. $N_\text{aff}$ is the number of big particles in region $v_\text{b}$ in the case that nonaffine displacement does not take place. It is given by $N_\text{aff}=\rho\int_{v_\text{b}}g_\text{aff}(\bm{r})\textrm{d}\bm{r}$. $p$ evaluates the probability that a particle in $v_\text{b}$ undergoes a significant nonaffine displacement in a lifecycle of the LER. Figure~\ref{fig:9}(a) shows the values of $p$ at all simulated conditions.

With above results and considerations, we can estimate the long-time nonaffine self-diffusivity of the big particle $D_\text{naff}$, which is defined by
\begin{equation}
D_\text{naff}=\lim_{\delta t\to\infty}\frac{\langle\tilde{\bm{r}}^2(\delta t)\rangle}{6\delta t}
,\label{eq:3_12}
\end{equation}
where $\tilde{\bm{r}}(\delta t)$ is the nonaffine displacement of a particle during a time interval $\delta t$ \cite{yama2}. In our picture, an LER sequentially experiences elastic deformation, yielding and flow in its lifecycle. The average time of the elastic deformation process is given by $\tau_\text{el}=2\bar\gamma/\dot\gamma$. The yielding and flow of LER are induced by large-step nonaffine displacement of particles. Assuming that such movements are realized by self-diffusion of particle, we find that the characteristic time of the yielding and flow, $\tau_\text{naff}$, is written as $\tau_\text{naff}\approx l_\text{s}^2/6D_0$. Then, the lifecycle of an LER is given by $\tau_\text{LER}\approx\tau_\text{el}+\tau_\text{naff}$. During one $\tau_\text{LER}$, some particles in the first shell undergo nonaffine large-step jump, while others' movements are restricted by their nearest neighbors. For the latter particles, their nonaffine mean square displacements, $\langle\tilde r_\text{slow}^2\rangle$, can be estimated by the plateau value in the double-logarithmic plot of $\langle\tilde{\bm{r}}^2(t)\rangle$. Noticing these two kinds of particle motion within one $\tau_\text{LER}$, the nonaffine mean square displacement of the particle in the first shell is estimated by $\langle\tilde{\bm{r}}_\text{LER}^2\rangle\approx p l_\text{s}^2+(1-p)\langle\tilde{r}_\text{slow}^2\rangle$. Here, we use $p$ defined in Eq.~(\ref{eq:3_11}) to approximate the probability that a particle in the first shell undergoes the nonaffine large-step jump. In principle, $p$ evaluates the particle jump in the first and third quadrants shown in Fig.~\ref{fig:7}(a), and the $\langle\tilde{\bm{r}}_\text{LER}^2\rangle$ found in this way is more suitable for characterizing the nonaffine motion along the extensional direction. Nevertheless, we can accept this rough approximation, because many previous studies suggest that the nonaffine mean square displacement does not strongly depend on the direction \cite{poon1, brady1, yama2}. With $\langle\tilde{\bm{r}}_\text{LER}^2\rangle$ and $\tau_\text{LER}$, we find a nonaffine diffusivity given by
\begin{equation}
D_\text{naff}^{\text{(LER)}}=\frac{\langle\tilde{\bm{r}}_\text{LER}^2\rangle}{6\tau_\text{LER}}
.\label{eq:3_13}
\end{equation}
We check the value of $\tau_\text{LER}$ for all simulated conditions. It is found that at $\delta t=\tau_\text{LER}$, $\langle\tilde{\bm{r}}^2(\delta t)\rangle$ almost attains the long-time diffusion behavior. Figure~\ref{fig:9}(b) displays the scatterplot of $D_\text{naff}$ and $D_\text{naff}^{\text{(LER)}}$ in the nonlinear regime. These two quantities exhibit very good linear correlation, as indicated by their Pearson correlation coefficient \cite{pearson} $\rho=0.998$. Performing a linear fit on the points shown in Fig.~\ref{fig:9}(b) results in a relation $D_\text{naff}=1.26D_\text{naff}^{\text{(LER)}}$. The agreement between $D_\text{naff}^{\text{(LER)}}$ and $D_\text{naff}$ is impressive, especially considering that they are obtained by different approaches. Their numerical difference could be due to the overestimation of $\tau_\text{naff}$. Here we calculate $\tau_\text{naff}$ with diffusive picture. While this process could be more ballistic \cite{pete2}.

\subsection{Discussion}

It is interesting to compare the LER with some existing approaches resolving the nonlinear rheology of glass and glass-forming liquids. In the study of nonlinear rheology of glass, the key question is to figure out \textit{why and how an amorphous solid flows}. The typical experimental setup is the startup shear \cite{weitz1, poon1, weeks1, pete2, pete4, egelhaaf2, schall5, fuchs1} (or startup extension for metallic glasses, see Refs.\cite{egami1, voigtmann2}). From the microscopic point of view, the concept of STZ is a natural choice for explaining the yielding of amorphous solids. An STZ is a liquid-like spot in the solid background, which plays as the precursor of the bulk yielding and flow. While for supercooled liquids, flowing is not a problem, and the key question is changed to \textit{why a liquid exhibits strong viscoelasticity}, such as the strong shear thinning. The steady shear, rather than the startup shear, is the typical setup for the study of flow behaviors of supercooled liquids \cite{hoover1, simmons1, miyazaki1, yama1, yama2, porcar1, wang3, rice1}. The change of the key question calls for a shift of consideration. One cannot just focus on the ``soft'' regions within which particles collectively undergo large nonaffine displacements, such as the STZ or the cooperatively rearranging region that we will discuss in the next paragraph. The shift of consideration leads to the concept of LER. In contrast to STZ, an LER is a solid-like spot in the liquid background. It provides the resistance to the imposed shear. Its deformation and rearrangement are the microscopic source of the nonlinear viscoelasticity of supercooled liquids. With the help of the elastic model of supercooled liquids \cite{dyre1, dyre2, dyre3, dyre4}, the connection between shear thinning and the evolution of LER is clearly established, as we have shown in sub-section III.A. Moreover, the LER picture offers a practical way to analyze the distortion of $g(\bm{r})$ by decomposing $g(\bm{r})$ into a strong, affine part $g_\text{aff}(\bm{r})$ and a smaller, nonaffine part $\Delta g(\bm{r})$. By analyzing $\Delta g(\bm{r})$, one can evaluate the nonaffine particle displacements hidden in the microscopic anisotropy, as we have presented in sub-section III.B. In many previous studies of glass rheology \cite{pete2, pete4, fuchs1, voigtmann2}, researchers adopt $\delta g(\bm{r})$, defined as $\delta g(\bm{r})=g(\bm{r})-g_\text{eq}(r)$, to represent the microscopic structural distortion. According to the LER picture, $\delta g(\bm{r})$ contains both the affine and nonaffine ingredients, and thus might bring ambiguity in its interpretation.

As mentioned in section I, many studies of the nonlinear rheology of supercooled liquids are based on the idea of dynamical heterogeneity. For example, Yamamoto and Onuki find that the bond, defined as the connectivity between two neighboring particles, breaks collectively in the form of cluster in the shear thinning regime of soft-repulsive supercooled liquids \cite{yama1, yama2}. Such cluster of bond breakage (CBB) is consistent with the celebrated concept of cooperatively rearranging region put forth by Adam and Gibbs \cite{adam1}. As expected, the length scale of CBB $\xi_\text{CBB}$ grows as the quiescent system approaches the glassy state. In sheared supercooled liquids, the dynamical heterogeneity and its associated CBBs are significant in Newtonian regime. In the shear thinning regime, the CBB shrinks with shear rate by $\xi_\text{CBB}\sim\dot\gamma^{-0.5}$ in the three-dimensional system \cite{yama2}. Outside the CBB, most particles still keep their connectivity with neighboring particles, and thus should move affinely in the flow. So, LER and CBB are the negative of each other to some extent. Both LER and CBB are temporary and fluctuating. It is possible that after the relaxation of an LER, the region transforms to a CBB. It is inspiring to find that in the Newtonian regime, the dynamical heterogeneity is strong, while the LER mechanism is not important. On the contrary, in the nonlinear regime, LER mechanism is dominant, while the dynamical heterogeneity is suppressed. Therefore, it seems that the smooth transition from the Newtonian regime to the nonlinear regime is accompanied by the competition between the dynamical heterogeneity effect and the LER mechanism. Our research on this problem is under progress. It should be pointed out that the concepts of CBB and LER are derived from very different considerations. Thus, we cannot expect them to exhibit same dependences on shear or the degree of supercooling. 

The MCT-ITT approach \cite{fuchs2, fuchs5}, especially its schematic form \cite{cates2}, highlights the caging effect in the nonlinear rheology of deeply supercooled liquid and glass, and works well for concentrated hard-sphere-like microgel suspensions \cite{fuchs3, fuchs4}. The concept of cage elasticity has also been employed to discuss the rheological data of hard-sphere glasses \cite{pete1, poon2, pete2, pete3, pete4}. Indeed, the nearest neighbors contribute most to the atomic level stress \cite{egami5} of the reference particle. Nevertheless, as indicated by Eqs.~(\ref{eq:3_7}) and (\ref{eq:3_8}), by considering the elasticity that extends beyond the spatial range of the cage, one establishes a straightforward connection between the elastic properties and the nonlinear rheology of supercooled liquids.

Notice that, there is no abrupt change in the structure going from a supercooled liquid to a glass, at least at the level of two-point correlation functions such as $g(\bm{r})$. From the viewpoint of dynamics, both supercooled liquids and glasses are featured by the significant limit on the diffusion of particle, which results in the emergence of elasticity. In the case that the flow is fast enough, the local structure will be relaxed mainly through the shear-driven process for both of deeply supercooled liquids and glasses. Considering these similarities, we suggest that the concept of LER should also be applicable to the shear thinning of glassy materials under steady homogeneous flow, though it is introduced based on supercooled liquids.

\begin{figure*}[htb]
\includegraphics[scale=1]{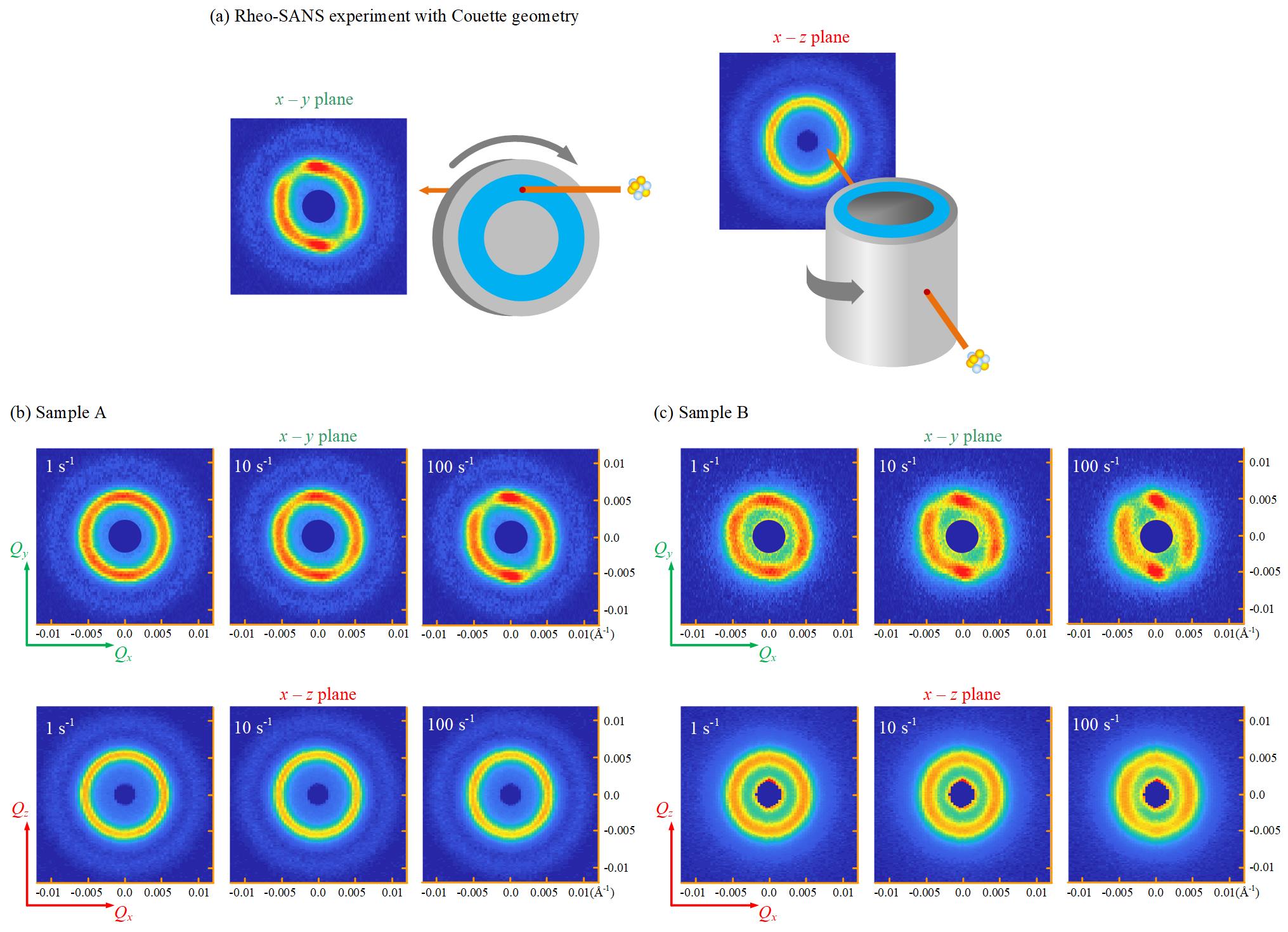}
\caption{(a) Illustration of the Rheo-SANS experiment under Couette geometry. $x$, $y$ and $z$ denote the directions of flow, velocity gradient and vorticity, respectively. (b) 2D SANS patterns obtained from the $x-y$ plane and the $x-z$ plane at $\dot\gamma=1$, $10$ and $100 \text{ s}^{-1}$ for sample A. (c) 2D SANS patterns obtained from the $x-y$ plane and the $x-z$ plane at $\dot\gamma=1$, $10$ and $100\text{ s}^{-1}$ for sample B. We did not see long-range ordering in all measured 2D patterns. The sample thicknesses along the neutron beam are 5 mm and 2 mm for the $x-y$ plane and the $x-z$ plane, respectively.
\label{fig:10}}
\end{figure*}

\section{RHEO-SANS EXPERIMENT}

As introduced in the beginning, confocal microscopy serves as an invaluable tool for revealing the structure and dynamics of colloidal glass at the particle level. However, owing to the technical limit, its use in suspensions subject to fast steady shear (e.g., $\dot\gamma\gg 1 \text{ s}^{-1}$) or bulk fluids with thickness larger than 1 mm could be restricted. On the other hand, neutron has the merit of strong penetrability \cite{squires1}, which allows the thickness of the sample to reach several millimeters. Moreover, neutron scattering measures the average structural factor of the sample via natural interference \cite{squires1}, and therefore is particularly suitable for the study of the suspensions in steady state. In the past two decades, Rheo-SANS technique has been extensively adopted to study the nonlinear rheology \cite{wagner1, wagner2, wagner3, wang3, watanabe1} and flow-induced ordering or melting \cite{vermant1, ackerson3} of colloidal suspensions under steady shear. With above considerations, we use Rheo-SANS technique with Couette geometry \cite{porcar1, burgh1} to experimentally explore the microscopic origin of the nonlinear viscoelasticity in the supercooled colloidal suspension. In passing, Rheo-Small Angle X-ray Scattering is another state-of-the-art scattering technique to investigate the nonlinear glassy rheology \cite{schall4, schall5, fuchs1}, which provides better resolution than Rheo-SANS, but is restricted by sample thickness.

\begin{figure*}
\includegraphics[scale=1]{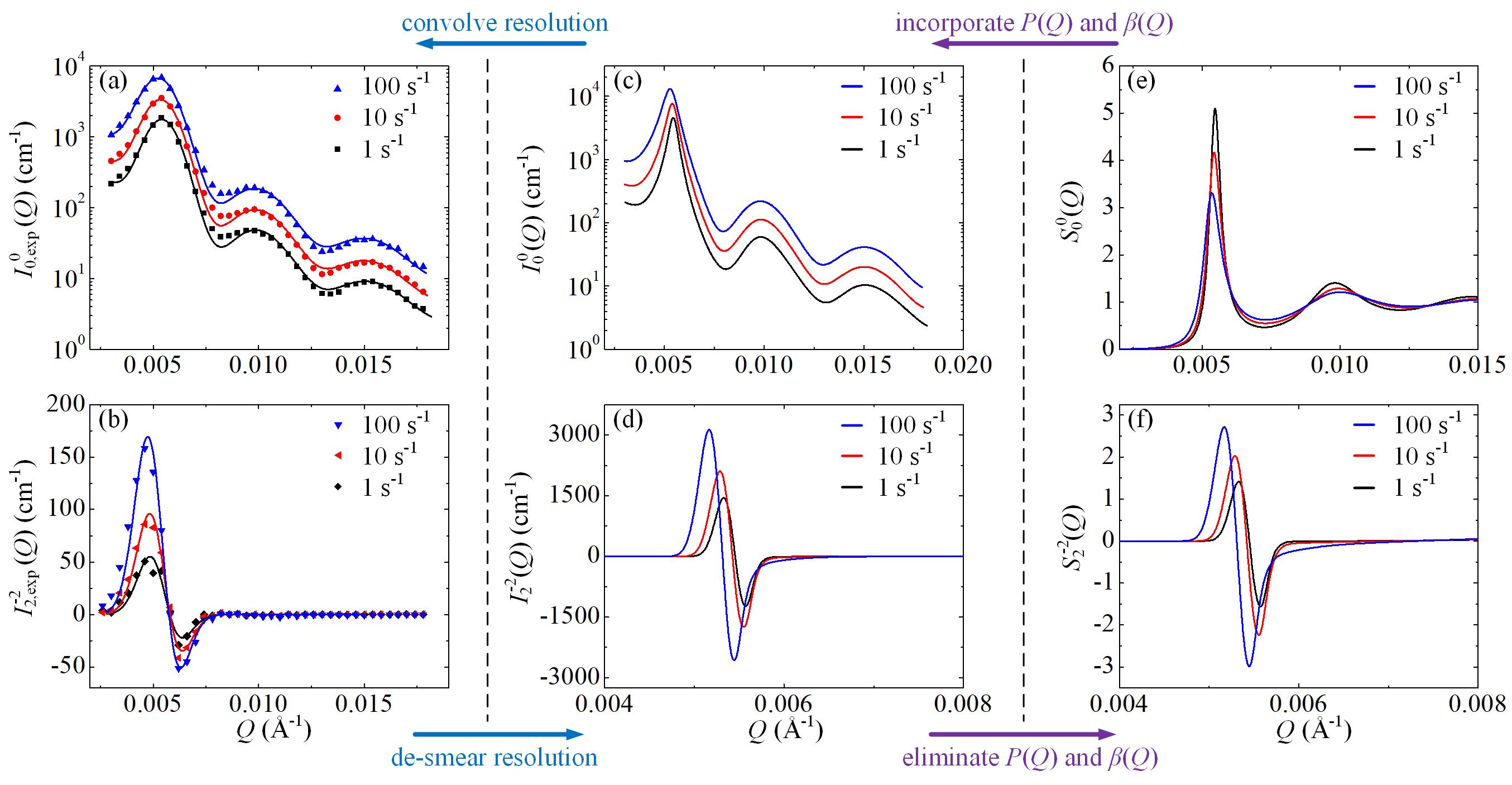}
\caption{Finding $S_0^0(Q)$ and $S_2^{-2}(Q)$ for sample A from the Rheo-SANS spectra. (a) and (b) respectively show $I_{0,\text{exp}}^0(Q)$ and $I_{2,\text{exp}}^{-2}(Q)$ at $\dot\gamma=1$, $10$, and $100\text{ s}^{-1}$ extracted from the 2D SANS patterns through the method given in Appendix B. Experimental data are denoted by symbols. Solid lines are calculated by convolving the $I_0^0(Q)$ and $I_2^{-2}(Q)$ given in panels (c) and (d) with instrumental resolution function. (c) and (d) respectively show $I_0^0(Q)$ and $I_2^{-2}(Q)$ obtained from $I_{0,\text{exp}}^0(Q)$ and $I_{2,\text{exp}}^{-2}(Q)$ by eliminating the resolution effect. (e) and (f) respectively show $S_0^0(Q)$ and $S_2^{-2}(Q)$ obtained from $I_0^0(Q)$ and $I_2^{-2}(Q)$ by eliminating the effects of $P(Q)$ and $\beta(Q)$. For clarity, we vertically shift the data shown in panels (a) and (c).
\label{fig:11}}
\end{figure*}

The Rheo-SANS experiment was performed at the D22 SANS beamline at the Institut Laue-Langevin. The measurements of the $x-y$ plane and the $x-z$ plane were respectively realized with a home-made flow cell and the Anton Paar MCR 501 rheometer. The shear viscosity of the suspension was measured during the Rheo-SANS experiment. Figure~\ref{fig:10} illustrates the experiment and displays some typical 2D SANS patterns in the two planes. We measured following two colloidal suspensions. Sample A is composed of charged silica particles suspended in a solvent consisting of a mixture of ethylene glycol and glycerol. The volume fraction of the silica particle is about 40\%. The Kob-Andersen mixture of two kinds of silica particles \cite{kob1}, with diameter of 120 nm and 80 nm in a number ratio of 4:1, was used to avoid shear-induced crystallization. Sample B is composed of charged silica particles suspended in the glycerol. The volume fraction of the silica particles is about 35\%. The particles possess an average diameter $D$ of 120 nm and a size polydispersity ($\sigma_D/D$, where $\sigma_D$ is the standard deviation of $D$) of 13\%. The strong polydispersity effectively prevents the sample from shear-induced long-range ordering. For both samples, the proton to deuterium ratio of the solvent was carefully adjusted to avoid multiple neutron scattering \cite{wang3}.

For polydisperse colloidal suspensions, the SANS spectrum $I(\bm{Q})$ can be treated by the $\beta$ approximation \cite{shc1}
\begin{equation}
I(\bm{Q})=n_\text{p}AP(Q)S'(\bm{Q})
,\label{eq:4_1}
\end{equation}
where $\bm{Q}$ is the scattering vector in scattering experiments, $n_\text{p}$ is the number density of colloids, $A$ denotes the contrast of the scattering length between solute particle and solvent, $P(Q)$ is the average form factor normalized at zero scattering angle, and $S'(\bm{Q})$ is the apparent structure factor given by $S'(\bm{Q})=1+\beta(Q)[S(\bm{Q})-1]$, where $\beta(Q)$ is the polydispersity factor \cite{shc1} and $S(\bm{Q})$ is the interparticle structure factor \cite{hansen1}. For our silica particles, $P(Q)$ can be well described by the spherical core-shell model \cite{shc2}. For sheared colloids, $I(\bm{Q})$ and $S(\bm{Q})$ are anisotropic, and can be expanded by spherical harmonics in the way similar to Eq.~(\ref{eq:2_3}). Then, we have
\begin{eqnarray}
I_0^0(Q)&=&n_\text{p}AP(Q)\lbrace 1+\beta(Q)[S_0^0(Q)-1]\rbrace
,\label{eq:4_2}\\
I_2^{-2}(Q)&=&n_\text{p}AP(Q)\beta(Q)S_2^{-2}(Q)
,\label{eq:4_3}
\end{eqnarray}
where $I_l^m(Q)$ and $S_l^m(Q)$ are expansion coefficients corresponding to $Y_l^m(\bm{Q}/Q)$. Experimental $I_0^0(Q)$ and $I_2^{-2}(Q)$, which we denote as $I_{0,\text{exp}}^0(Q)$ and $I_{2,\text{exp}}^{-2}(Q)$, can be obtained from the measured 2D SANS patterns in the $x-y$ and $x-z$ planes. The detail in extracting $I_{0,\text{exp}}^0(Q)$ and $I_{2,\text{exp}}^{-2}(Q)$ is given in Appendix B. Note that, the measured spectra are smeared by the instrumental resolution as
\begin{equation}
I_{l,\text{exp}}^m(Q)=I_l^m(Q)\otimes R(Q)
,\label{eq:4_4}
\end{equation}
where $R(Q)$ is the instrumental resolution function, and $\otimes$ denotes the convolution. Therefore, to obtain $S_0^0(Q)$ and $S_2^{-2}(Q)$ from $I_{0,\text{exp}}^0(Q)$ and $I_{2,\text{exp}}^{-2}(Q)$ with Eqs.~(\ref{eq:4_2}) and (\ref{eq:4_3}), one needs to first de-smear the instrumental resolution from the measured spectra, and then eliminate the influences of $P(Q)$ and $\beta(Q)$. Taking sample A as an example, Fig.~\ref{fig:11} gives a typical procedure of finding $S_0^0(Q)$ and $S_2^{-2}(Q)$ from experiments. As seen from Fig.~\ref{fig:11}(b) and (f), microscopic anisotropy is noticeably enhanced with shear.

\begin{figure}
\includegraphics[scale=1]{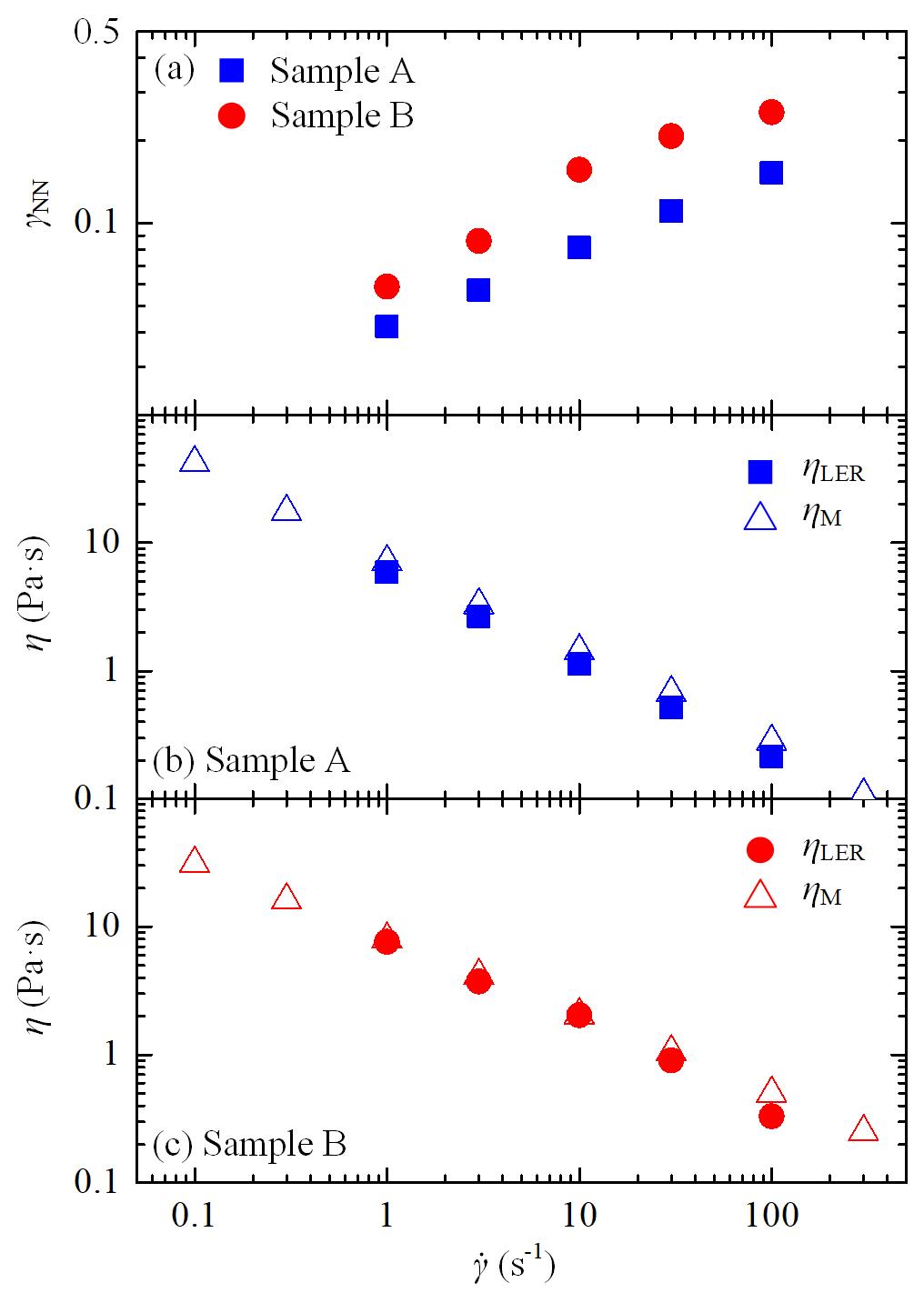}
\caption{(a) Experimental $\gamma_\text{NN}$ found through Eq.~(\ref{eq:4_9}) for both samples. (b) Comparison between the microscopic shear viscosity $\eta_\text{LER}$ and the macroscopic shear viscosity $\eta_\text{M}$ for sample A. (c) Comparison between $\eta_\text{LER}$ and $\eta_\text{M}$ for sample B.
\label{fig:12}}
\end{figure}

The analyses shown in section II and III are carried out in real space. While scattering experiment measures the reciprocal space. $S(\bm{Q})$ and $g(\bm{r})$ form a Fourier pair \cite{hansen1}. Their SHE coefficients are related by the spherical Bessel transformation \cite{egami3}
\begin{eqnarray}
g_l^m(r)&=&\frac{\mathrm{i}^l}{2\pi^2n_\text{p}}\int S_l^m(Q)j_l(Qr)Q^2\textrm{d}Q
,\label{eq:4_5}\\
S_l^m(Q)&=&\mathrm{i}^l4\pi n_\text{p}\int g_l^m(r)j_l(Qr)r^2\textrm{d}r
,\label{eq:4_6}
\end{eqnarray}
where $j_l(x)$ is the $l$-order spherical Bessel function of the first kind. In the $Q$-space, the distortion for an affine shear with a small strain $\gamma$ is expressed as
\begin{equation}
S_2^{-2}(Q)=\gamma S_\text{iso}(Q)=\frac{\gamma}{\sqrt{15}}Q\frac{\textrm{d}}{\textrm{d}Q}S_0^0(Q)
,\label{eq:4_7}
\end{equation}
where
\begin{equation}
S_\text{iso}(Q)=\frac{1}{\sqrt{15}}Q\frac{\textrm{d}}{\textrm{d}Q}S_0^0(Q)
.\label{eq:4_8}
\end{equation}
Equation~(\ref{eq:4_7}) has a form similar to Eq.~(\ref{eq:2_7}). According to Eq.~(\ref{eq:4_7}), we can find the average strain of LER $\bar\gamma$ by minimizing $L=\sum_j[S_2^{-2}(Q_j)-\bar\gamma S_\text{iso}(Q_j)]^2$ where $Q_j$ denotes the measured $Q$ values. Nevertheless, as suggested by Eq.~(\ref{eq:3_5}) and Fig.~\ref{fig:3}, the microscopic stress is mainly determined by the distortion of the cage $\gamma_{\text{NN}}$ rather than $\bar\gamma$. Notice that, many studies have established the correlation between the cage configuration and the intensity and anisotropy of the main peak of $S(\bm{Q})$ \cite{schall4, schall5, fuchs1, hansen1}. Therefore, $\gamma_{\text{NN}}$ could be quantitatively related to the first positive peaks of $S_2^{-2}(Q)$ and $S_0^0(Q)$. We explore this possible relation with our BD results. With Eq.~(\ref{eq:4_6}), we calculate $S_2^{-2}(Q)$ and $S_0^0(Q)$ from the simulated $g(\bm{r})$, and find that $\gamma_{\text{NN}}$ can be nicely estimated by
\begin{equation}
\gamma_{\text{NN}}\approx\frac{A_{1\text{p}}[S_2^{-2}(Q)]}{A_{1\text{p}}[S_\text{iso}(Q)]}
,\label{eq:4_9}
\end{equation}
where $A_{1\text{p}}[f(Q)]$ denotes the area of the first positive peak of $f(Q)$. The results of $\gamma_\text{NN}$ found from this empirical approach for both samples are shown in Fig.~\ref{fig:12}(a). As volume fraction increases, $\gamma_\text{NN}$ is seen to decrease and become less sensitive to shear rate. These observations are consistent with our simulation results and the evolution of the 2D anisotropic patterns shown in Fig.~\ref{fig:10}(b) and (c). With $\gamma_\text{NN}$, we compute the microscopic shear viscosity $\eta_\text{LER}=G_\infty\gamma_\text{NN}/\dot\gamma$, where $G_\infty$ is estimated by the storage modulus $G'$ found in the small-angle oscillatory shear measurement \cite{wang3, rogers1}. The results are displayed in Fig.~\ref{fig:12}(b) and (c). We also show the macroscopic shear viscosity $\eta_\text{M}$ of both samples in Fig.~\ref{fig:12}(b) and (c). Here, $\eta_\text{M}$ is given by $\eta_\text{M}=\eta_\text{tot}-\eta_\text{sol}(1+2.5\phi)$, where $\eta_\text{tot}$ is the total shear viscosity of sample measured by rheometer, and $\eta_\text{sol}$ is the viscosity of solvent. $\eta_\text{M}$ exhibits a power-law dependence on $\dot\gamma$, suggesting that the flows are in the nonlinear regime of glassy liquids. As seen from Fig.~\ref{fig:12}(b) and (c), the agreement between $\eta_\text{LER}$ and $\eta_\text{M}$ is remarkable. It confirms our prediction that localized elasticity governs the shear thinning of supercooled and glassy liquids.

\begin{figure}
\includegraphics[scale=1]{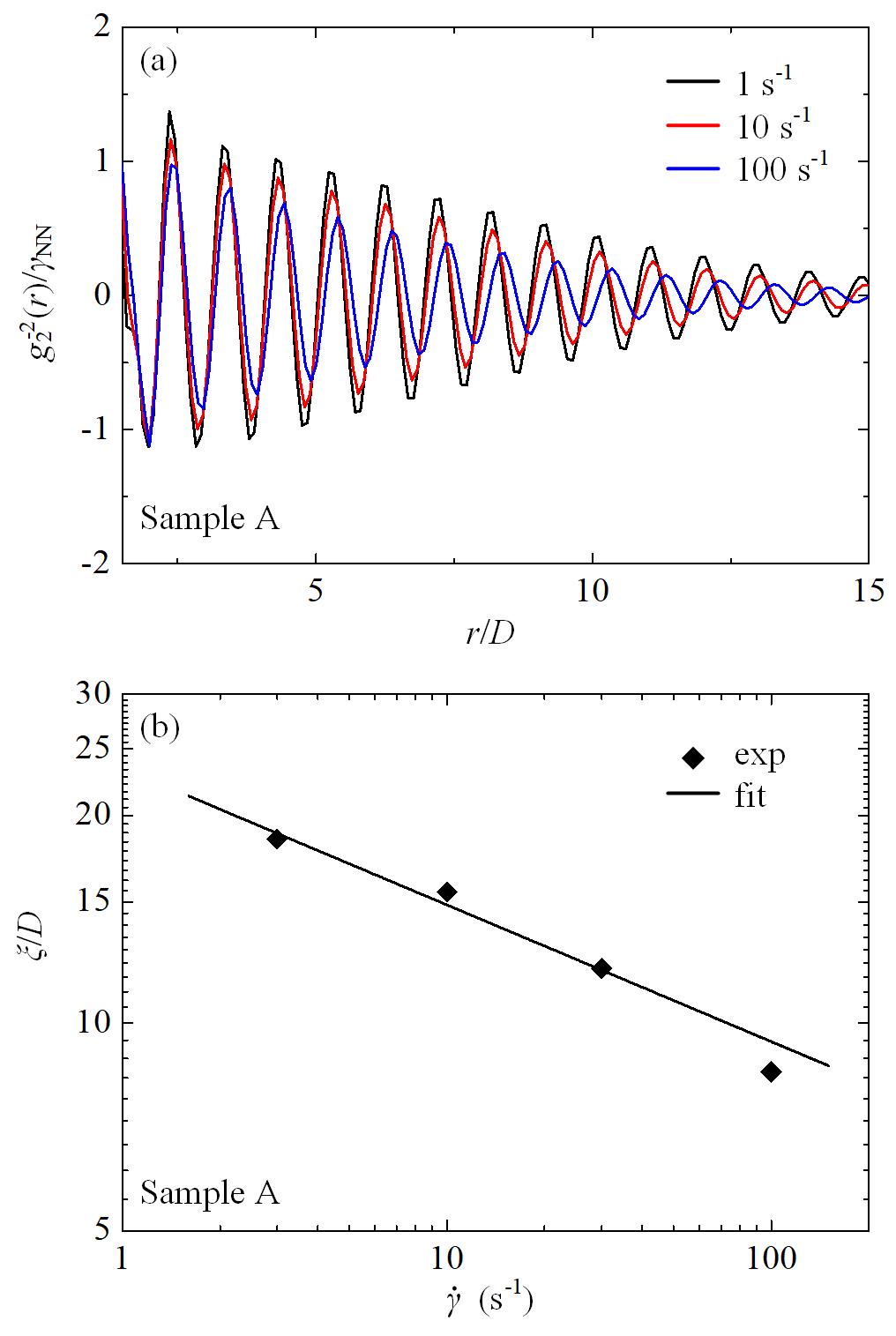}
\caption{(a) Experimental $g_2^{-2}(r)$ of sample A at $\dot\gamma=1$, $10$, and $100\text{ s}^{-1}$. The magnitude of $g_2^{-2}(r)$ is normalized by $\gamma_\text{NN}$. (b) The range of LER $\xi$ (symbols) of sample A estimated from experimental $g_2^{-2}(r)$. Here, we do not plot the result at $\dot\gamma=1\text{ s}^{-1}$ since its value is unreasonably large. The solid line denotes the fit with the power law $\xi\propto\dot\gamma^{-\nu}$, where $\nu$ is found to be 0.198.
\label{fig:13}}
\end{figure}

With Eq.~(\ref{eq:4_5}), we can calculate $g_0^0(r)$ and $g_2^{-2}(r)$ from $S_0^0(Q)$ and $S_2^{-2}(Q)$, and use them to evaluate the spatial range of the LER $\xi$.  Figure~\ref{fig:13}(a) gives some examples of $g_2^{-2}(r)$ of sample A. Here, we do not show the curves at $r<2D$ ($D=120$ nm is the diameter of big particle in sample A), because the results at small $r$ are severely deteriorated by the cut-off error in experimental $S_2^{-2}(Q)$\footnote{This is why we cannot find $\gamma_\text{NN}$ with experimental $g_2^{-2}(r)$. $\gamma_\text{NN}$ is determined by $g_2^{-2}(r)$ at small $r$, which is severely deteriorated by the cut-off error in experimental $S_2^{-2}(Q)$.}. As expected, the amplitude of $g_2^{-2}(r)$ decays as $r$ increases. Such decay becomes more evident as shear rate increases, suggesting the shrinkage of LER with increasing shear rate. Figure~\ref{fig:13}(b) shows the results of $\xi$ of sample A. We respectively fit the experimental $\eta_\text{M}(\dot\gamma)$, $\gamma_\text{NN}(\dot\gamma)$, and $\xi(\dot\gamma)$ of sample A with $\eta_\text{M}\propto\dot\gamma^{-\lambda}$, $\gamma_\text{NN}\propto\dot\gamma^\epsilon$ and $\xi\propto\dot\gamma^{-\nu}$, as what we did in section III. The exponents are found to be $\lambda=0.728$, $\epsilon=0.281$, and $\nu=0.198$. Therefore, we have $1-\epsilon=0.719$ and $4\nu=0.792$. The values of $\lambda$, $1-\epsilon$, and $4\nu$ are reasonably close to each other.

Numerically, the shrinkage of the spatial range of $g_2^{-2}(r)$ with shear rate is due to the broadening of the first positive and the first negative peaks of $S_2^{-2}(Q)$. The values of $\xi$ found from experiment depend on the details in the extraction of $S_2^{-2}(Q)$ from $I_{2,\text{exp}}^{-2}(Q)$. For example, in our analysis, we use the Bessel function of the first kind to describe the first positive peak of $S_2^{-2}(Q)$. By replacing it with a broader function, the values of experimental $\xi$ can be smaller. Nevertheless, we emphasize that the shrinkage of $\xi$ with shear rate does not depend on the specific strategy in data analysis. In fact, the broadening of the first positive peak of $S_2^{-2}(Q)$ with shear rate can be directly observed from the raw data.

\section{CONCLUSION}

In summary, we connect the shear thinning of colloidal supercooled liquids to the transient elastic behavior in the flow by introducing the concept of localized elastic region. LER is the microscopic structural unit that provides the resistance to the imposed shear. Upon increasing the shear rate $\dot\gamma$, the size of LER shrinks as $\dot\gamma^{-\nu}$, while its characteristic strain increases as $\dot\gamma^\epsilon$. Three exponents, $\nu$, $\epsilon$, and $\lambda$ that describes the extent of shear thinning by $\eta\sim\dot\gamma^{-\lambda}$, are related by $\lambda=4\nu=1-\epsilon$. This equation, which is a natural derivation of the LER picture, connects the bulk nonlinear viscoelasticity to the microscopic configurational distortion and elastic properties of the system. The relaxation of LER is mainly realized by the large-step nonaffine particle displacement along the extensional direction of the shear geometry. Such effect grows with shear rate, which contributes to the enhancement of nonaffine diffusion as shear rate increases. These results offer a new perspective for understanding the nonlinear rheology and viscoelasticity of glass-forming liquids with long-range repulsive interactions.

\begin{acknowledgments}
This research was supported by the National Natural Science Foundation of China (Grant No. 11975136). Part of this research was performed at the Spallation Neutron Source, which is US Department of Energy (DOE) Office of Science User Facilities operated by Oak Ridge National Laboratory. We are grateful to the D22 beamline at Institut Laue-Langevin for the Rheo-SANS measurement.
\end{acknowledgments}

\appendix
\section{CAGE JUMP}

In this part, we calculate the cage jump length $l_\text{cj}$ following the method given by Candelier \textit{et al}. \cite{biroli1, biroli2}. First, it splits the trajectory of a particle $S(t)_{t\in[0,T]}$ into two sets of successive points, respectively denoted as $S_1$ and $S_2$, at an arbitrary cut time $t_\text{c}$. One can evaluate how well separated are these two sets of points by
\begin{equation}
p(t_\text{c})=\zeta(t_\text{c})[\langle d_1^2(t_2)\rangle_{t_2\in S_2}\langle d_2^2(t_1)\rangle_{t_1\in S_1}]^{1/2}
,\label{eq:A_1}
\end{equation}
where $\zeta(t_\text{c})=\sqrt{t_\text{c}/T\times(1-t_\text{c}/T)}$, $d_k(t_i)$ s the distance between the point at time $t_i$ and the center of mass of the subset $S_k$. The average $\langle\dots\rangle_{S_k}$ is over the subset $S_k$. A cage jump event is defined at $t_\text{c}$ if $p(t_\text{c})$ is maximal. Then, by iteratively repeating this procedure for every sub-trajectory until $p_\text{max}(t_\text{c})$ is smaller than a threshold $\sigma_\text{c}^2$ ($\sigma_\text{c}$ is the cage size), one separates the total trajectory into caging motions connected by jumps. $\sigma_\text{c}^2$ is determined as the crossover from the plateau behavior to the long-time diffusive behavior in the double-logarithmic plot of the particle mean square displacement as a function of time. It is found to be $0.05d_\text{b}^2$ in our case.

\begin{figure}
\includegraphics[scale=1]{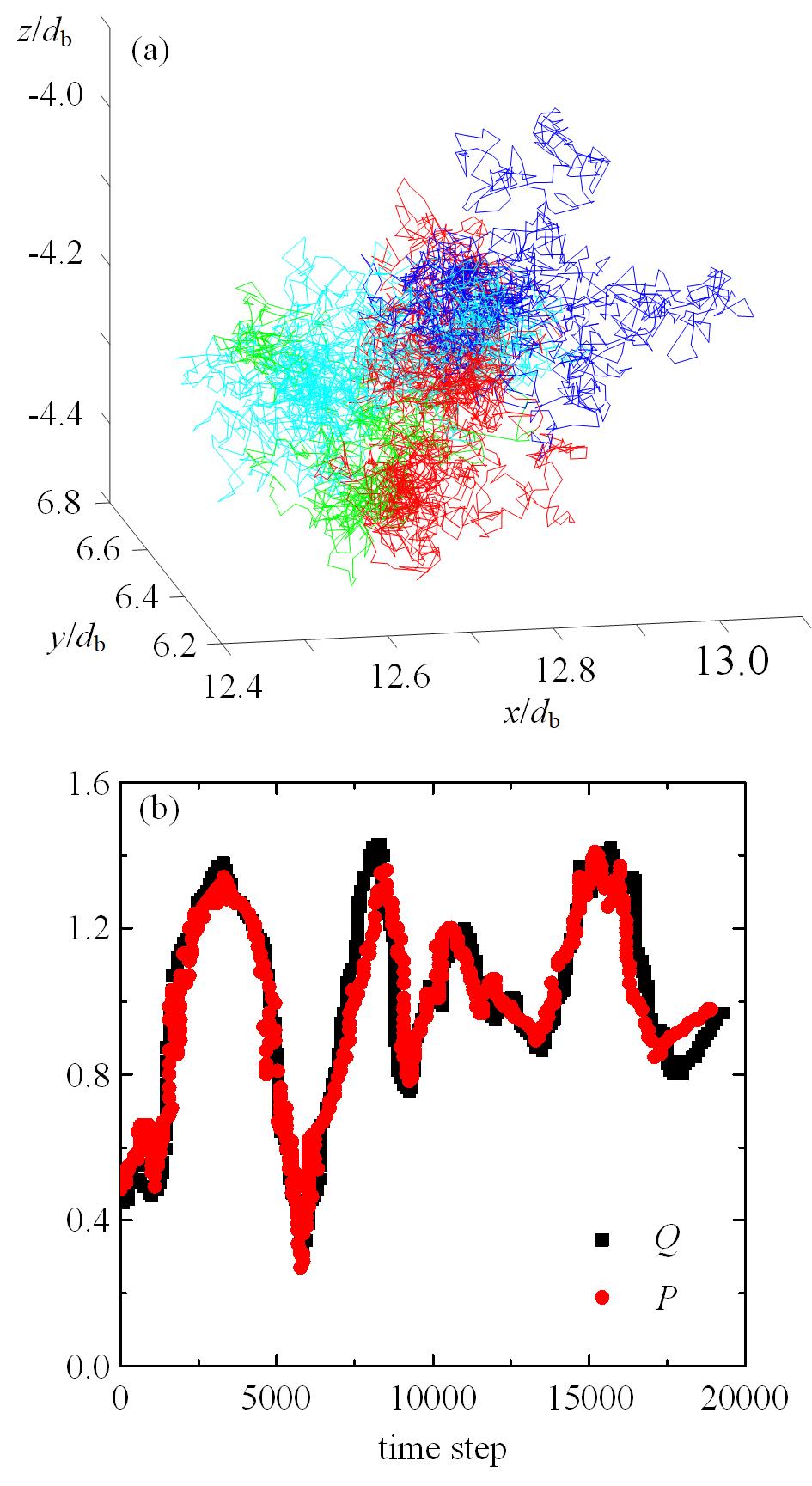}
\caption{(a) Trajectory of a reference particle. The color of trajectory changes at every jump. There are 4 segments shown here. (b) Comparison between $Q_t(a^\ast,\tau^\ast)/\langle Q_t\rangle_t$ and $P_t(\tau^\ast)/\langle P_t\rangle_t$. The simulation condition here is $\phi=45\%$ and $\dot\gamma=0$.
\label{fig:A1}}
\end{figure}

We applied this method on big particles at $\phi=45\%$ and $\dot\gamma=0$. $l_\text{cj}$ is found to be $0.418d_\text{b}$. Figure~\ref{fig:A1}(a) illustrates the separation of the particle trajectory by cage jump. We verify our analysis with the four-point correlation function \cite{glotzer1}, as suggested by Candelier \textit{et al}. \cite{biroli1, biroli2}. For particle $j$, the following function can be calculated
\begin{equation}
Q_{j,t}(a,\tau)=\exp\Big[-\frac{|\Delta\bm{r}_j(t,t+\tau)|^2}{2a^2}\Big]
,\label{eq:A_2}
\end{equation}
where $\Delta\bm{r}_j(t,t+\tau)$ is the displacement of particle $j$ between $t$ and $t+\tau$, $a$ is a probing length scale. With $Q_{j,t}(a,\tau)$, we calculate the four-point correlation function $\chi_4(a,\tau)$
\begin{equation}
\chi_4(a,\tau)=N[\langle Q_t^2(a,\tau)\rangle-\langle Q_t(a,\tau)\rangle^2]
,\label{eq:A_3}
\end{equation}
where $N$ is the particle number, and $Q_t(a,\tau)$ is defined as $Q_t(a,\tau)=\sum_jQ_{j,t}(a,\tau)/N$. $\chi_4(a,\tau)$ reaches maximal at $\tau^\ast$ and $a^\ast$, indicating a dynamical heterogeneity. We compute $Q_t(a^\ast,\tau^\ast)/\langle Q_t\rangle_t$ and the relative percentage $P_t(\tau^\ast)/\langle P_t\rangle_t$ of particles that have not jumped between $t$ and $t+\tau^\ast$. These two quantities are compared in Fig.~\ref{fig:A1}(b), and seen to match each other well. Note that, $Q_t(a^\ast,\tau^\ast)/\langle Q_t\rangle_t$ measures the “immobile” particles. Thus, the agreement between these two quantities supports our analysis on the cage jump.

To find the cage jump length in flowing state, we adopt the nonaffine particle displacement $\tilde{\bm{r}}(t)$ defined by Yamamoto and Onuki \cite{yama2} as the particle trajectory. The results for the $\phi=45\%$ system at $\textrm{Pe}=0.25$, $0.75$, $2.5$ and $5$ are respectively 0.411, 0.402, 0.394 and 0.386 ($d_\text{b}$). These values are slightly smaller than that in the zero shear, and decreases with shear rate, as expected.

\section{FINDING \boldmath{$I_{0,\text{exp}}^0(Q)$} AND \boldmath{$I_{2,\text{exp}}^{-2}(Q)$}}

In this part, we give a description of finding $I_{0,\text{exp}}^0(Q)$ and $I_{2,\text{exp}}^{-2}(Q)$ from the Rheo-SANS experiment under the Couette geometry. In following paragraphs, we will drop the subscript “exp” for shortness’ sake.

For anisotropic system, the SANS pattern can be expanded by the real spherical harmonics
\begin{equation}
I(\bm{Q})=\sum_{l=0}^\infty\sum_{m=-l}^l I_l^m(Q)Y_l^m\Big(\frac{\bm{Q}}{Q}\Big)
.\label{eq:A_4}
\end{equation}
The real spherical harmonics are mutually orthogonal in three-dimensional space
\begin{equation}
\int\textrm{d}\bm{\Omega}Y_l^m(\bm{\Omega})Y_{l'}^{m'}(\bm{\Omega})=4\pi\delta_{ll'}\delta_{mm'}
.\label{eq:A_5}
\end{equation}
Due to the symmetry of Couette geometry, in Eq.~(\ref{eq:A_4}), only terms with even $l$ and $m$ survives.

In experiment, we can only access $I(\bm{Q})$ in two planes. Thus, to find $I_l^m(Q)$, Eq.~(\ref{eq:A_4}) needs to be simplified. According to the analysis in II. Theoretical Framework, terms with $l\ge4$ are of higher orders of $\gamma$. In the case that $\gamma$ is small, we can tentatively simplify Eq.~(\ref{eq:A_4}) by ignoring these terms
\begin{equation}
\begin{aligned}
I(\bm{Q})&\approx I_0^0(Q)Y_0^0\Big(\frac{\bm{Q}}{Q}\Big)+I_2^{-2}(Q)Y_2^{-2}\Big(\frac{\bm{Q}}{Q}\Big)\\
&+I_2^0(Q)Y_2^0\Big(\frac{\bm{Q}}{Q}\Big)+I_2^2(Q)Y_2^2\Big(\frac{\bm{Q}}{Q}\Big)
.\label{eq:A_6}
\end{aligned}
\end{equation}
Notice that, in the nonlinear regime, $S_4^4(Q)$ can also be significant. Therefore, the validity of above approximation should be inspected. We will do it in the end of this part.

\begin{figure}
\includegraphics[scale=1]{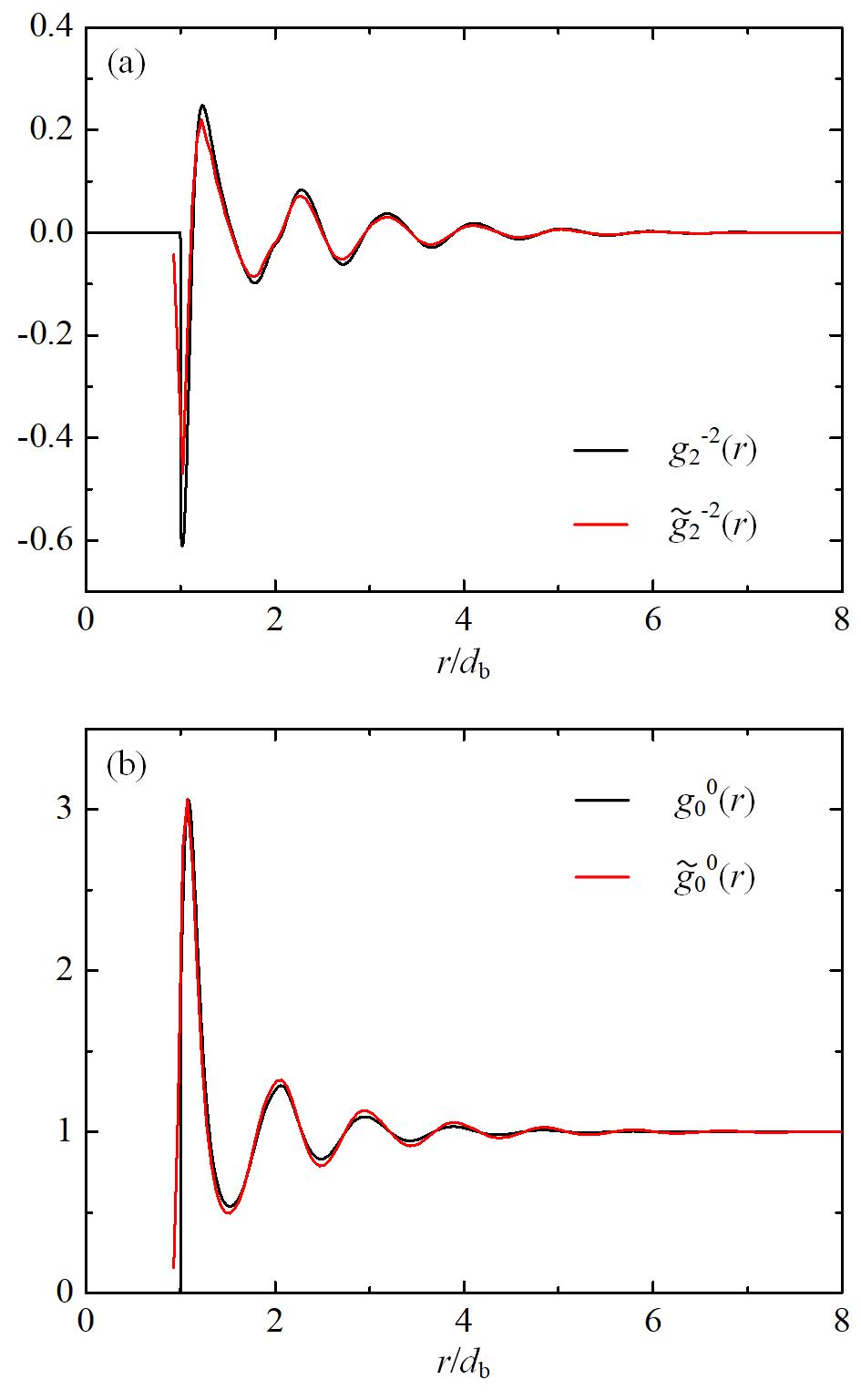}
\caption{(a) Comparison between $\tilde g_2^{-2}(r)$ and $g_2^{-2}(r)$. (b) Comparison between $\tilde g_0^0(r)$ and $g_0^0(r)$. The simulation condition is $\phi=45\%$, $\textrm{Pe}=0.75$.
\label{fig:A2}}
\end{figure}

From the measured pattern in $x-y$ plane ($I(Q,\theta=\pi/2,\phi)$), we can calculate the following quantities
\begin{equation}
\begin{aligned}
I_{2,-2}^{xy}(Q)&=\frac{1}{2\pi}\int_0^{2\pi}I\Big(Q,\theta=\frac{\pi}{2},\phi\Big)Y_2^{-2}\Big(\theta=\frac{\pi}{2},\phi\Big)\textrm{d}\phi\\
&=\frac{1}{2\pi}\int_0^{2\pi}I\Big(Q,\theta=\frac{\pi}{2},\phi\Big)\frac{\sqrt{15}}{2}\sin{2\phi}\textrm{d}\phi
,\label{eq:A_7}
\end{aligned}
\end{equation}
\begin{equation}
\begin{aligned}
I_{2,2}^{xy}(Q)&=\frac{1}{2\pi}\int_0^{2\pi}I\Big(Q,\theta=\frac{\pi}{2},\phi\Big)Y_2^{2}\Big(\theta=\frac{\pi}{2},\phi\Big)\textrm{d}\phi\\
&=\frac{1}{2\pi}\int_0^{2\pi}I\Big(Q,\theta=\frac{\pi}{2},\phi\Big)\frac{\sqrt{15}}{2}\cos{2\phi}\textrm{d}\phi
.\label{eq:A_8}
\end{aligned}
\end{equation}
From the measured pattern in $x-z$ plane ($I(Q,\theta,\phi=0)$), the following quantity can be found
\begin{equation}
\begin{aligned}
I_{0,0}^{xz}(Q)&=\frac{1}{2}\int_0^{\pi}I(Q,\theta,\phi=0)Y_0^0\sin\theta\textrm{d}\theta\\
&=\frac{1}{2}\int_0^{\pi}I(Q,\theta,\phi=0)\sin\theta\textrm{d}\theta
.\label{eq:A_9}
\end{aligned}
\end{equation}
Combining Eqs.~(\ref{eq:A_5}-\ref{eq:A_9}), it is straightforward to show that
\begin{eqnarray}
I_{2,-2}^{xy}(Q)&=&\frac{15}{8}I_2^{-2}(Q)
,\label{eq:A_10}\\
I_{2,2}^{xy}(Q)&=&\frac{15}{8}I_2^2(Q)
,\label{eq:A_11}\\
I_{0,0}^{xz}(Q)&=&I_0^0(Q)+\sqrt{\frac{5}{3}}I_2^2(Q)
.\label{eq:A_12}
\end{eqnarray}
Then, we have
\begin{eqnarray}
I_2^{-2}(Q)&=&\frac{8}{15}I_{2,-2}^{xy}(Q)
,\label{eq:A_13}\\
I_0^0(Q)&=&I_{0,0}^{xz}(Q)-\frac{8}{3\sqrt{15}}I_{2,2}^{xy}(Q)
.\label{eq:A_14}
\end{eqnarray}
With the preceding two equations, we can obtain $I_0^0(Q)$ and $I_2^{-2}(Q)$ from the Rheo-SANS experiment.

We test the validity of the above approximation with our BD results at $\phi=45\%$, $\textrm{Pe}=0.75$, which well locates in the shear thinning regime. We first generate the 2D cross-section of $g(\bm{r})$ in the $x-y$ and $x-z$ planes with the thickness of the plane to be $0.8d_\text{b}$. Then, we calculate $g_0^0(r)$ and $g_2^{-2}(r)$ with the approximation given by Eqs.~(\ref{eq:A_13}) and (\ref{eq:A_14})
\begin{eqnarray}
\tilde g_2^{-2}(r)&=&\frac{8}{15}g_{2,-2}^{xy}(r)
,\label{eq:A_15}\\
\tilde g_0^0(r)&=&g_{0,0}^{xz}(r)-\frac{8}{3\sqrt{15}}g_{2,2}^{xy}(r)
,\label{eq:A_16}
\end{eqnarray}
where $\tilde{}$ denotes the approximated results. We compare $\tilde g_2^{-2}(r)$ and $\tilde g_0^0(r)$ with $g_2^{-2}(r)$ and $g_0^0(r)$ in Fig.~\ref{fig:A2}. It is seen that the above approximation is acceptable. We find that $\tilde g_2^2(r)$ remarkably deviates from $g_2^2(r)$. Fortunately, our analysis does not involve this term.

%\nocite{*}
\bibliography{LER2022bib}

\end{document}